\pdfoutput=1
\documentclass[11pt,a4paper]{article}

\usepackage[T1]{fontenc}
\usepackage[utf8]{inputenc}
\usepackage{graphicx}
\usepackage{amsmath}
\usepackage{amssymb}
\usepackage[table]{xcolor}
\usepackage{booktabs}
\usepackage{tabularx}
\usepackage[margin=1in]{geometry}
\usepackage{float}
\usepackage{url}
\expandafter\def\expandafter\UrlBreaks\expandafter{\UrlBreaks\do\-}
\usepackage{hyperref}
\usepackage{tikz}
\usepackage{pgfplots}
\pgfplotsset{compat=1.18}
\usetikzlibrary{arrows.meta,positioning}

\usepackage[skip=6pt]{caption}
\usepackage{pgfplots}
\pgfplotsset{compat=1.18}
\usetikzlibrary{arrows.meta,positioning}

\tikzset{
  agent/.style={draw, rounded corners=2pt, minimum width=1.55cm, minimum height=0.44cm, font=\footnotesize, inner sep=1pt},
  callbox/.style={agent, minimum height=0.9cm},
  floorbar/.style={draw, fill=black!6, rounded corners=1.5pt, minimum width=0.3cm, font=\footnotesize},
  conv/.style={agent, minimum width=1.2cm},
  flow/.style={-{Latex[length=1.6mm]}, thin},
  floorflow/.style={{Latex[length=1.4mm]}-{Latex[length=1.4mm]}, thin, black!55},
  mlab/.style={font=\tiny\itshape, text=black!65},
  hdr/.style={font=\footnotesize\bfseries},
}

\newif\ifdraftnotes
\draftnotesfalse

\newcommand{\res}[1]{\expandafter\ifx\csname res:#1\endcsname\relax\textbf{\textcolor{red}{??}}\else\csname res:#1\endcsname\fi}
\newcommand{\restable}[1]{\csname res:table:#1\endcsname}
\expandafter\def\csname res:table:main:function\endcsname{\multicolumn{6}{c}{\textcolor{red}{results pending}}\\}
\expandafter\def\csname res:table:main:class\endcsname{\multicolumn{6}{c}{\textcolor{red}{results pending}}\\}
\expandafter\def\csname res:table:pairs:all\endcsname{\multicolumn{7}{c}{\textcolor{red}{results pending}}\\}
\expandafter\def\csname res:runs\endcsname{2430}
\expandafter\def\csname res:tasks\endcsname{135}
\expandafter\def\csname res:tasks:function\endcsname{50}
\expandafter\def\csname res:tasks:class\endcsname{85}
\expandafter\def\csname res:reps\endcsname{3}
\expandafter\def\csname res:spend\endcsname{67.16}
\expandafter\def\csname res:date\endcsname{2026-10-15}
\expandafter\def\csname res:dateend\endcsname{2026-10-16}
\expandafter\def\csname res:strong\endcsname{claude-opus-5-5}
\expandafter\def\csname res:cheap\endcsname{claude-haiku-5-5}
\expandafter\def\csname res:price:strong\endcsname{\$4/\$20}
\expandafter\def\csname res:price:cheap\endcsname{\$0.1/\$0.5}
\expandafter\def\csname res:r\endcsname{0.025}
\expandafter\def\csname res:rin\endcsname{0.025}
\expandafter\def\csname res:ceiling\endcsname{40}
\expandafter\def\csname res:pass:function:A\endcsname{100.0}
\expandafter\def\csname res:passlo:function:A\endcsname{100.0}
\expandafter\def\csname res:passhi:function:A\endcsname{100.0}
\expandafter\def\csname res:cps:function:A\endcsname{0.0063}
\expandafter\def\csname res:cost:function:A\endcsname{0.0056}
\expandafter\def\csname res:wall:function:A\endcsname{2.8}
\expandafter\def\csname res:in:function:A\endcsname{274}
\expandafter\def\csname res:out:function:A\endcsname{261}
\expandafter\def\csname res:think:function:A\endcsname{16}
\expandafter\def\csname res:thinkshare:function:A\endcsname{6.0}
\expandafter\def\csname res:pass:function:Aprime\endcsname{98.7}
\expandafter\def\csname res:passlo:function:Aprime\endcsname{96.0}
\expandafter\def\csname res:passhi:function:Aprime\endcsname{100.0}
\expandafter\def\csname res:cps:function:Aprime\endcsname{0.00019}
\expandafter\def\csname res:cost:function:Aprime\endcsname{0.00014}
\expandafter\def\csname res:wall:function:Aprime\endcsname{1.7}
\expandafter\def\csname res:in:function:Aprime\endcsname{274}
\expandafter\def\csname res:out:function:Aprime\endcsname{323}
\expandafter\def\csname res:think:function:Aprime\endcsname{125}
\expandafter\def\csname res:thinkshare:function:Aprime\endcsname{38.7}
\expandafter\def\csname res:pass:function:R\endcsname{98.7}
\expandafter\def\csname res:passlo:function:R\endcsname{96.7}
\expandafter\def\csname res:passhi:function:R\endcsname{100.0}
\expandafter\def\csname res:cps:function:R\endcsname{0.00045}
\expandafter\def\csname res:cost:function:R\endcsname{0.00022}
\expandafter\def\csname res:wall:function:R\endcsname{2.7}
\expandafter\def\csname res:in:function:R\endcsname{609}
\expandafter\def\csname res:out:function:R\endcsname{396}
\expandafter\def\csname res:think:function:R\endcsname{183}
\expandafter\def\csname res:thinkshare:function:R\endcsname{46.3}
\expandafter\def\csname res:pass:function:B\endcsname{100.0}
\expandafter\def\csname res:passlo:function:B\endcsname{100.0}
\expandafter\def\csname res:passhi:function:B\endcsname{100.0}
\expandafter\def\csname res:cps:function:B\endcsname{0.0279}
\expandafter\def\csname res:cost:function:B\endcsname{0.0272}
\expandafter\def\csname res:wall:function:B\endcsname{12.8}
\expandafter\def\csname res:in:function:B\endcsname{1521}
\expandafter\def\csname res:out:function:B\endcsname{1091}
\expandafter\def\csname res:think:function:B\endcsname{212}
\expandafter\def\csname res:thinkshare:function:B\endcsname{19.4}
\expandafter\def\csname res:pass:function:C\endcsname{100.0}
\expandafter\def\csname res:passlo:function:C\endcsname{100.0}
\expandafter\def\csname res:passhi:function:C\endcsname{100.0}
\expandafter\def\csname res:cps:function:C\endcsname{0.0279}
\expandafter\def\csname res:cost:function:C\endcsname{0.0270}
\expandafter\def\csname res:wall:function:C\endcsname{13.0}
\expandafter\def\csname res:in:function:C\endcsname{1522}
\expandafter\def\csname res:out:function:C\endcsname{1090}
\expandafter\def\csname res:think:function:C\endcsname{215}
\expandafter\def\csname res:thinkshare:function:C\endcsname{19.7}
\expandafter\def\csname res:pass:function:D\endcsname{100.0}
\expandafter\def\csname res:passlo:function:D\endcsname{100.0}
\expandafter\def\csname res:passhi:function:D\endcsname{100.0}
\expandafter\def\csname res:cps:function:D\endcsname{0.0074}
\expandafter\def\csname res:cost:function:D\endcsname{0.0068}
\expandafter\def\csname res:wall:function:D\endcsname{9.1}
\expandafter\def\csname res:in:function:D\endcsname{1322}
\expandafter\def\csname res:out:function:D\endcsname{1380}
\expandafter\def\csname res:think:function:D\endcsname{708}
\expandafter\def\csname res:thinkshare:function:D\endcsname{51.3}
\expandafter\def\csname res:table:main:function\endcsname{A: single call, strong & 100.0 & 0.0063 & 0.0056 & 2.8 & 274 / 261 (16) \\
A$'$: single call, cheap & 98.7 & 0.00019 & 0.00014 & 1.7 & 274 / 323 (125) \\
R: router & 98.7 & 0.00045 & 0.00022 & 2.7 & 609 / 396 (183) \\
B: multi-call, plain & 100.0 & 0.0279 & 0.0272 & 12.8 & 1521 / 1091 (212) \\
C: multi-call, OFP uniform & 100.0 & 0.0279 & 0.0270 & 13.0 & 1522 / 1090 (215) \\
D: multi-call, OFP mixed & 100.0 & 0.0074 & 0.0068 & 9.1 & 1322 / 1380 (708) \\}
\expandafter\def\csname res:pass:class:A\endcsname{90.2}
\expandafter\def\csname res:passlo:class:A\endcsname{84.3}
\expandafter\def\csname res:passhi:class:A\endcsname{95.3}
\expandafter\def\csname res:cps:class:A\endcsname{0.0287}
\expandafter\def\csname res:cost:class:A\endcsname{0.0244}
\expandafter\def\csname res:wall:class:A\endcsname{8.4}
\expandafter\def\csname res:in:class:A\endcsname{893}
\expandafter\def\csname res:out:class:A\endcsname{1114}
\expandafter\def\csname res:think:class:A\endcsname{244}
\expandafter\def\csname res:thinkshare:class:A\endcsname{21.9}
\expandafter\def\csname res:pass:class:Aprime\endcsname{52.5}
\expandafter\def\csname res:passlo:class:Aprime\endcsname{42.0}
\expandafter\def\csname res:passhi:class:Aprime\endcsname{62.7}
\expandafter\def\csname res:cps:class:Aprime\endcsname{0.0020}
\expandafter\def\csname res:cost:class:Aprime\endcsname{0.00093}
\expandafter\def\csname res:wall:class:Aprime\endcsname{6.4}
\expandafter\def\csname res:in:class:Aprime\endcsname{893}
\expandafter\def\csname res:out:class:Aprime\endcsname{1942}
\expandafter\def\csname res:think:class:Aprime\endcsname{771}
\expandafter\def\csname res:thinkshare:class:Aprime\endcsname{39.7}
\expandafter\def\csname res:pass:class:R\endcsname{55.3}
\expandafter\def\csname res:passlo:class:R\endcsname{45.5}
\expandafter\def\csname res:passhi:class:R\endcsname{65.5}
\expandafter\def\csname res:cps:class:R\endcsname{0.0090}
\expandafter\def\csname res:cost:class:R\endcsname{0.0011}
\expandafter\def\csname res:wall:class:R\endcsname{7.7}
\expandafter\def\csname res:in:class:R\endcsname{1847}
\expandafter\def\csname res:out:class:R\endcsname{1860}
\expandafter\def\csname res:think:class:R\endcsname{725}
\expandafter\def\csname res:thinkshare:class:R\endcsname{39.0}
\expandafter\def\csname res:pass:class:B\endcsname{89.4}
\expandafter\def\csname res:passlo:class:B\endcsname{83.1}
\expandafter\def\csname res:passhi:class:B\endcsname{94.9}
\expandafter\def\csname res:cps:class:B\endcsname{0.0875}
\expandafter\def\csname res:cost:class:B\endcsname{0.0740}
\expandafter\def\csname res:wall:class:B\endcsname{26.2}
\expandafter\def\csname res:in:class:B\endcsname{4257}
\expandafter\def\csname res:out:class:B\endcsname{3060}
\expandafter\def\csname res:think:class:B\endcsname{666}
\expandafter\def\csname res:thinkshare:class:B\endcsname{21.8}
\expandafter\def\csname res:pass:class:C\endcsname{87.8}
\expandafter\def\csname res:passlo:class:C\endcsname{81.2}
\expandafter\def\csname res:passhi:class:C\endcsname{93.7}
\expandafter\def\csname res:cps:class:C\endcsname{0.0898}
\expandafter\def\csname res:cost:class:C\endcsname{0.0754}
\expandafter\def\csname res:wall:class:C\endcsname{26.6}
\expandafter\def\csname res:in:class:C\endcsname{4263}
\expandafter\def\csname res:out:class:C\endcsname{3092}
\expandafter\def\csname res:think:class:C\endcsname{693}
\expandafter\def\csname res:thinkshare:class:C\endcsname{22.4}
\expandafter\def\csname res:pass:class:D\endcsname{77.3}
\expandafter\def\csname res:passlo:class:D\endcsname{69.0}
\expandafter\def\csname res:passhi:class:D\endcsname{85.1}
\expandafter\def\csname res:cps:class:D\endcsname{0.0429}
\expandafter\def\csname res:cost:class:D\endcsname{0.0305}
\expandafter\def\csname res:wall:class:D\endcsname{20.8}
\expandafter\def\csname res:in:class:D\endcsname{4633}
\expandafter\def\csname res:out:class:D\endcsname{4381}
\expandafter\def\csname res:think:class:D\endcsname{1536}
\expandafter\def\csname res:thinkshare:class:D\endcsname{35.1}
\expandafter\def\csname res:table:main:class\endcsname{A: single call, strong & 90.2 & 0.0287 & 0.0244 & 8.4 & 893 / 1114 (244) \\
A$'$: single call, cheap & 52.5 & 0.0020 & 0.00093 & 6.4 & 893 / 1942 (771) \\
R: router & 55.3 & 0.0090 & 0.0011 & 7.7 & 1847 / 1860 (725) \\
B: multi-call, plain & 89.4 & 0.0875 & 0.0740 & 26.2 & 4257 / 3060 (666) \\
C: multi-call, OFP uniform & 87.8 & 0.0898 & 0.0754 & 26.6 & 4263 / 3092 (693) \\
D: multi-call, OFP mixed & 77.3 & 0.0429 & 0.0305 & 20.8 & 4633 / 4381 (1536) \\}
\expandafter\def\csname res:pass:all:A\endcsname{93.8}
\expandafter\def\csname res:passlo:all:A\endcsname{90.1}
\expandafter\def\csname res:passhi:all:A\endcsname{97.0}
\expandafter\def\csname res:cps:all:A\endcsname{0.0198}
\expandafter\def\csname res:cost:all:A\endcsname{0.0166}
\expandafter\def\csname res:wall:all:A\endcsname{6.1}
\expandafter\def\csname res:in:all:A\endcsname{664}
\expandafter\def\csname res:out:all:A\endcsname{798}
\expandafter\def\csname res:think:all:A\endcsname{160}
\expandafter\def\csname res:thinkshare:all:A\endcsname{20.0}
\expandafter\def\csname res:pass:all:Aprime\endcsname{69.6}
\expandafter\def\csname res:passlo:all:Aprime\endcsname{62.0}
\expandafter\def\csname res:passhi:all:Aprime\endcsname{77.0}
\expandafter\def\csname res:cps:all:Aprime\endcsname{0.0011}
\expandafter\def\csname res:cost:all:Aprime\endcsname{0.00066}
\expandafter\def\csname res:wall:all:Aprime\endcsname{4.6}
\expandafter\def\csname res:in:all:Aprime\endcsname{664}
\expandafter\def\csname res:out:all:Aprime\endcsname{1342}
\expandafter\def\csname res:think:all:Aprime\endcsname{532}
\expandafter\def\csname res:thinkshare:all:Aprime\endcsname{39.6}
\expandafter\def\csname res:pass:all:R\endcsname{71.4}
\expandafter\def\csname res:passlo:all:R\endcsname{64.0}
\expandafter\def\csname res:passhi:all:R\endcsname{78.5}
\expandafter\def\csname res:cps:all:R\endcsname{0.0046}
\expandafter\def\csname res:cost:all:R\endcsname{0.00080}
\expandafter\def\csname res:wall:all:R\endcsname{6.0}
\expandafter\def\csname res:in:all:R\endcsname{1389}
\expandafter\def\csname res:out:all:R\endcsname{1318}
\expandafter\def\csname res:think:all:R\endcsname{524}
\expandafter\def\csname res:thinkshare:all:R\endcsname{39.8}
\expandafter\def\csname res:pass:all:B\endcsname{93.3}
\expandafter\def\csname res:passlo:all:B\endcsname{89.1}
\expandafter\def\csname res:passhi:all:B\endcsname{96.8}
\expandafter\def\csname res:cps:all:B\endcsname{0.0638}
\expandafter\def\csname res:cost:all:B\endcsname{0.0567}
\expandafter\def\csname res:wall:all:B\endcsname{20.5}
\expandafter\def\csname res:in:all:B\endcsname{3243}
\expandafter\def\csname res:out:all:B\endcsname{2331}
\expandafter\def\csname res:think:all:B\endcsname{498}
\expandafter\def\csname res:thinkshare:all:B\endcsname{21.4}
\expandafter\def\csname res:pass:all:C\endcsname{92.3}
\expandafter\def\csname res:passlo:all:C\endcsname{88.1}
\expandafter\def\csname res:passhi:all:C\endcsname{96.0}
\expandafter\def\csname res:cps:all:C\endcsname{0.0650}
\expandafter\def\csname res:cost:all:C\endcsname{0.0586}
\expandafter\def\csname res:wall:all:C\endcsname{21.1}
\expandafter\def\csname res:in:all:C\endcsname{3248}
\expandafter\def\csname res:out:all:C\endcsname{2350}
\expandafter\def\csname res:think:all:C\endcsname{516}
\expandafter\def\csname res:thinkshare:all:C\endcsname{21.9}
\expandafter\def\csname res:pass:all:D\endcsname{85.7}
\expandafter\def\csname res:passlo:all:D\endcsname{80.0}
\expandafter\def\csname res:passhi:all:D\endcsname{90.9}
\expandafter\def\csname res:cps:all:D\endcsname{0.0275}
\expandafter\def\csname res:cost:all:D\endcsname{0.0226}
\expandafter\def\csname res:wall:all:D\endcsname{16.6}
\expandafter\def\csname res:in:all:D\endcsname{3407}
\expandafter\def\csname res:out:all:D\endcsname{3270}
\expandafter\def\csname res:think:all:D\endcsname{1229}
\expandafter\def\csname res:thinkshare:all:D\endcsname{37.6}
\expandafter\def\csname res:table:main:all\endcsname{A: single call, strong & 93.8 & 0.0198 & 0.0166 & 6.1 & 664 / 798 (160) \\
A$'$: single call, cheap & 69.6 & 0.0011 & 0.00066 & 4.6 & 664 / 1342 (532) \\
R: router & 71.4 & 0.0046 & 0.00080 & 6.0 & 1389 / 1318 (524) \\
B: multi-call, plain & 93.3 & 0.0638 & 0.0567 & 20.5 & 3243 / 2331 (498) \\
C: multi-call, OFP uniform & 92.3 & 0.0650 & 0.0586 & 21.1 & 3248 / 2350 (516) \\
D: multi-call, OFP mixed & 85.7 & 0.0275 & 0.0226 & 16.6 & 3407 / 3270 (1229) \\}
\expandafter\def\csname res:costratio:function:B:A\endcsname{4.42}
\expandafter\def\csname res:costratiolo:function:B:A\endcsname{4.09}
\expandafter\def\csname res:costratiohi:function:B:A\endcsname{4.79}
\expandafter\def\csname res:costp:function:B:A\endcsname{\ensuremath{<10^{-4}}}
\expandafter\def\csname res:wallratio:function:B:A\endcsname{4.40}
\expandafter\def\csname res:wallratiolo:function:B:A\endcsname{4.15}
\expandafter\def\csname res:wallratiohi:function:B:A\endcsname{4.68}
\expandafter\def\csname res:wallp:function:B:A\endcsname{\ensuremath{<10^{-4}}}
\expandafter\def\csname res:passdiff:function:B:A\endcsname{+0.0}
\expandafter\def\csname res:passdifflo:function:B:A\endcsname{+0.0}
\expandafter\def\csname res:passdiffhi:function:B:A\endcsname{+0.0}
\expandafter\def\csname res:passp:function:B:A\endcsname{1.00}
\expandafter\def\csname res:costratio:function:C:A\endcsname{4.42}
\expandafter\def\csname res:costratiolo:function:C:A\endcsname{4.11}
\expandafter\def\csname res:costratiohi:function:C:A\endcsname{4.77}
\expandafter\def\csname res:costp:function:C:A\endcsname{\ensuremath{<10^{-4}}}
\expandafter\def\csname res:wallratio:function:C:A\endcsname{4.41}
\expandafter\def\csname res:wallratiolo:function:C:A\endcsname{4.16}
\expandafter\def\csname res:wallratiohi:function:C:A\endcsname{4.68}
\expandafter\def\csname res:wallp:function:C:A\endcsname{\ensuremath{<10^{-4}}}
\expandafter\def\csname res:passdiff:function:C:A\endcsname{+0.0}
\expandafter\def\csname res:passdifflo:function:C:A\endcsname{+0.0}
\expandafter\def\csname res:passdiffhi:function:C:A\endcsname{+0.0}
\expandafter\def\csname res:passp:function:C:A\endcsname{1.00}
\expandafter\def\csname res:costratio:function:C:B\endcsname{1.00}
\expandafter\def\csname res:costratiolo:function:C:B\endcsname{0.98}
\expandafter\def\csname res:costratiohi:function:C:B\endcsname{1.02}
\expandafter\def\csname res:costp:function:C:B\endcsname{0.86}
\expandafter\def\csname res:wallratio:function:C:B\endcsname{1.00}
\expandafter\def\csname res:wallratiolo:function:C:B\endcsname{0.98}
\expandafter\def\csname res:wallratiohi:function:C:B\endcsname{1.03}
\expandafter\def\csname res:wallp:function:C:B\endcsname{0.92}
\expandafter\def\csname res:passdiff:function:C:B\endcsname{+0.0}
\expandafter\def\csname res:passdifflo:function:C:B\endcsname{+0.0}
\expandafter\def\csname res:passdiffhi:function:C:B\endcsname{+0.0}
\expandafter\def\csname res:passp:function:C:B\endcsname{1.00}
\expandafter\def\csname res:costratio:function:D:C\endcsname{0.27}
\expandafter\def\csname res:costratiolo:function:D:C\endcsname{0.25}
\expandafter\def\csname res:costratiohi:function:D:C\endcsname{0.28}
\expandafter\def\csname res:costp:function:D:C\endcsname{\ensuremath{<10^{-4}}}
\expandafter\def\csname res:wallratio:function:D:C\endcsname{0.70}
\expandafter\def\csname res:wallratiolo:function:D:C\endcsname{0.67}
\expandafter\def\csname res:wallratiohi:function:D:C\endcsname{0.74}
\expandafter\def\csname res:wallp:function:D:C\endcsname{\ensuremath{<10^{-4}}}
\expandafter\def\csname res:passdiff:function:D:C\endcsname{+0.0}
\expandafter\def\csname res:passdifflo:function:D:C\endcsname{+0.0}
\expandafter\def\csname res:passdiffhi:function:D:C\endcsname{+0.0}
\expandafter\def\csname res:passp:function:D:C\endcsname{1.00}
\expandafter\def\csname res:costratio:function:D:A\endcsname{1.17}
\expandafter\def\csname res:costratiolo:function:D:A\endcsname{1.10}
\expandafter\def\csname res:costratiohi:function:D:A\endcsname{1.26}
\expandafter\def\csname res:costp:function:D:A\endcsname{\ensuremath{<10^{-4}}}
\expandafter\def\csname res:wallratio:function:D:A\endcsname{3.10}
\expandafter\def\csname res:wallratiolo:function:D:A\endcsname{2.88}
\expandafter\def\csname res:wallratiohi:function:D:A\endcsname{3.32}
\expandafter\def\csname res:wallp:function:D:A\endcsname{\ensuremath{<10^{-4}}}
\expandafter\def\csname res:passdiff:function:D:A\endcsname{+0.0}
\expandafter\def\csname res:passdifflo:function:D:A\endcsname{+0.0}
\expandafter\def\csname res:passdiffhi:function:D:A\endcsname{+0.0}
\expandafter\def\csname res:passp:function:D:A\endcsname{1.00}
\expandafter\def\csname res:costratio:function:Aprime:A\endcsname{0.03}
\expandafter\def\csname res:costratiolo:function:Aprime:A\endcsname{0.03}
\expandafter\def\csname res:costratiohi:function:Aprime:A\endcsname{0.03}
\expandafter\def\csname res:costp:function:Aprime:A\endcsname{\ensuremath{<10^{-4}}}
\expandafter\def\csname res:wallratio:function:Aprime:A\endcsname{0.66}
\expandafter\def\csname res:wallratiolo:function:Aprime:A\endcsname{0.57}
\expandafter\def\csname res:wallratiohi:function:Aprime:A\endcsname{0.75}
\expandafter\def\csname res:wallp:function:Aprime:A\endcsname{\ensuremath{<10^{-4}}}
\expandafter\def\csname res:passdiff:function:Aprime:A\endcsname{-1.3}
\expandafter\def\csname res:passdifflo:function:Aprime:A\endcsname{-4.0}
\expandafter\def\csname res:passdiffhi:function:Aprime:A\endcsname{+0.0}
\expandafter\def\csname res:passp:function:Aprime:A\endcsname{0.32}
\expandafter\def\csname res:costratio:function:R:A\endcsname{0.07}
\expandafter\def\csname res:costratiolo:function:R:A\endcsname{0.04}
\expandafter\def\csname res:costratiohi:function:R:A\endcsname{0.12}
\expandafter\def\csname res:costp:function:R:A\endcsname{\ensuremath{<10^{-4}}}
\expandafter\def\csname res:wallratio:function:R:A\endcsname{1.04}
\expandafter\def\csname res:wallratiolo:function:R:A\endcsname{0.94}
\expandafter\def\csname res:wallratiohi:function:R:A\endcsname{1.16}
\expandafter\def\csname res:wallp:function:R:A\endcsname{0.86}
\expandafter\def\csname res:passdiff:function:R:A\endcsname{-1.3}
\expandafter\def\csname res:passdifflo:function:R:A\endcsname{-3.3}
\expandafter\def\csname res:passdiffhi:function:R:A\endcsname{+0.0}
\expandafter\def\csname res:passp:function:R:A\endcsname{0.16}
\expandafter\def\csname res:costratio:function:D:Aprime\endcsname{39.2}
\expandafter\def\csname res:costratiolo:function:D:Aprime\endcsname{34.3}
\expandafter\def\csname res:costratiohi:function:D:Aprime\endcsname{45.1}
\expandafter\def\csname res:costp:function:D:Aprime\endcsname{\ensuremath{<10^{-4}}}
\expandafter\def\csname res:wallratio:function:D:Aprime\endcsname{4.70}
\expandafter\def\csname res:wallratiolo:function:D:Aprime\endcsname{4.18}
\expandafter\def\csname res:wallratiohi:function:D:Aprime\endcsname{5.32}
\expandafter\def\csname res:wallp:function:D:Aprime\endcsname{\ensuremath{<10^{-4}}}
\expandafter\def\csname res:passdiff:function:D:Aprime\endcsname{+1.3}
\expandafter\def\csname res:passdifflo:function:D:Aprime\endcsname{+0.0}
\expandafter\def\csname res:passdiffhi:function:D:Aprime\endcsname{+4.0}
\expandafter\def\csname res:passp:function:D:Aprime\endcsname{0.32}
\expandafter\def\csname res:costratio:function:R:D\endcsname{0.06}
\expandafter\def\csname res:costratiolo:function:R:D\endcsname{0.03}
\expandafter\def\csname res:costratiohi:function:R:D\endcsname{0.11}
\expandafter\def\csname res:costp:function:R:D\endcsname{\ensuremath{<10^{-4}}}
\expandafter\def\csname res:wallratio:function:R:D\endcsname{0.34}
\expandafter\def\csname res:wallratiolo:function:R:D\endcsname{0.31}
\expandafter\def\csname res:wallratiohi:function:R:D\endcsname{0.37}
\expandafter\def\csname res:wallp:function:R:D\endcsname{\ensuremath{<10^{-4}}}
\expandafter\def\csname res:passdiff:function:R:D\endcsname{-1.3}
\expandafter\def\csname res:passdifflo:function:R:D\endcsname{-3.3}
\expandafter\def\csname res:passdiffhi:function:R:D\endcsname{+0.0}
\expandafter\def\csname res:passp:function:R:D\endcsname{0.16}
\expandafter\def\csname res:table:pairs:function\endcsname{B vs A & 4.42 [4.09, 4.79] & \ensuremath{<10^{-4}} & 4.40 [4.15, 4.68] & \ensuremath{<10^{-4}} & +0.0 [+0.0, +0.0] & 1.00 \\
C vs A & 4.42 [4.11, 4.77] & \ensuremath{<10^{-4}} & 4.41 [4.16, 4.68] & \ensuremath{<10^{-4}} & +0.0 [+0.0, +0.0] & 1.00 \\
C vs B & 1.00 [0.98, 1.02] & 0.86 & 1.00 [0.98, 1.03] & 0.92 & +0.0 [+0.0, +0.0] & 1.00 \\
D vs C & 0.27 [0.25, 0.28] & \ensuremath{<10^{-4}} & 0.70 [0.67, 0.74] & \ensuremath{<10^{-4}} & +0.0 [+0.0, +0.0] & 1.00 \\
D vs A & 1.17 [1.10, 1.26] & \ensuremath{<10^{-4}} & 3.10 [2.88, 3.32] & \ensuremath{<10^{-4}} & +0.0 [+0.0, +0.0] & 1.00 \\
A$'$ vs A & 0.03 [0.03, 0.03] & \ensuremath{<10^{-4}} & 0.66 [0.57, 0.75] & \ensuremath{<10^{-4}} & -1.3 [-4.0, +0.0] & 0.32 \\
R vs A & 0.07 [0.04, 0.12] & \ensuremath{<10^{-4}} & 1.04 [0.94, 1.16] & 0.86 & -1.3 [-3.3, +0.0] & 0.16 \\
D vs A$'$ & 39.2 [34.3, 45.1] & \ensuremath{<10^{-4}} & 4.70 [4.18, 5.32] & \ensuremath{<10^{-4}} & +1.3 [+0.0, +4.0] & 0.32 \\
R vs D & 0.06 [0.03, 0.11] & \ensuremath{<10^{-4}} & 0.34 [0.31, 0.37] & \ensuremath{<10^{-4}} & -1.3 [-3.3, +0.0] & 0.16 \\}
\expandafter\def\csname res:costratio:class:B:A\endcsname{3.03}
\expandafter\def\csname res:costratiolo:class:B:A\endcsname{2.90}
\expandafter\def\csname res:costratiohi:class:B:A\endcsname{3.16}
\expandafter\def\csname res:costp:class:B:A\endcsname{\ensuremath{<10^{-4}}}
\expandafter\def\csname res:wallratio:class:B:A\endcsname{2.96}
\expandafter\def\csname res:wallratiolo:class:B:A\endcsname{2.83}
\expandafter\def\csname res:wallratiohi:class:B:A\endcsname{3.10}
\expandafter\def\csname res:wallp:class:B:A\endcsname{\ensuremath{<10^{-4}}}
\expandafter\def\csname res:passdiff:class:B:A\endcsname{-0.8}
\expandafter\def\csname res:passdifflo:class:B:A\endcsname{-5.1}
\expandafter\def\csname res:passdiffhi:class:B:A\endcsname{+3.9}
\expandafter\def\csname res:passp:class:B:A\endcsname{0.33}
\expandafter\def\csname res:costratio:class:C:A\endcsname{3.05}
\expandafter\def\csname res:costratiolo:class:C:A\endcsname{2.93}
\expandafter\def\csname res:costratiohi:class:C:A\endcsname{3.19}
\expandafter\def\csname res:costp:class:C:A\endcsname{\ensuremath{<10^{-4}}}
\expandafter\def\csname res:wallratio:class:C:A\endcsname{3.01}
\expandafter\def\csname res:wallratiolo:class:C:A\endcsname{2.88}
\expandafter\def\csname res:wallratiohi:class:C:A\endcsname{3.15}
\expandafter\def\csname res:wallp:class:C:A\endcsname{\ensuremath{<10^{-4}}}
\expandafter\def\csname res:passdiff:class:C:A\endcsname{-2.4}
\expandafter\def\csname res:passdifflo:class:C:A\endcsname{-6.3}
\expandafter\def\csname res:passdiffhi:class:C:A\endcsname{+1.2}
\expandafter\def\csname res:passp:class:C:A\endcsname{0.28}
\expandafter\def\csname res:costratio:class:C:B\endcsname{1.01}
\expandafter\def\csname res:costratiolo:class:C:B\endcsname{0.99}
\expandafter\def\csname res:costratiohi:class:C:B\endcsname{1.02}
\expandafter\def\csname res:costp:class:C:B\endcsname{0.29}
\expandafter\def\csname res:wallratio:class:C:B\endcsname{1.02}
\expandafter\def\csname res:wallratiolo:class:C:B\endcsname{1.00}
\expandafter\def\csname res:wallratiohi:class:C:B\endcsname{1.03}
\expandafter\def\csname res:wallp:class:C:B\endcsname{0.068}
\expandafter\def\csname res:passdiff:class:C:B\endcsname{-1.6}
\expandafter\def\csname res:passdifflo:class:C:B\endcsname{-5.5}
\expandafter\def\csname res:passdiffhi:class:C:B\endcsname{+2.4}
\expandafter\def\csname res:passp:class:C:B\endcsname{0.38}
\expandafter\def\csname res:costratio:class:D:C\endcsname{0.42}
\expandafter\def\csname res:costratiolo:class:D:C\endcsname{0.40}
\expandafter\def\csname res:costratiohi:class:D:C\endcsname{0.44}
\expandafter\def\csname res:costp:class:D:C\endcsname{\ensuremath{<10^{-4}}}
\expandafter\def\csname res:wallratio:class:D:C\endcsname{0.83}
\expandafter\def\csname res:wallratiolo:class:D:C\endcsname{0.79}
\expandafter\def\csname res:wallratiohi:class:D:C\endcsname{0.87}
\expandafter\def\csname res:wallp:class:D:C\endcsname{\ensuremath{<10^{-4}}}
\expandafter\def\csname res:passdiff:class:D:C\endcsname{-10.6}
\expandafter\def\csname res:passdifflo:class:D:C\endcsname{-16.9}
\expandafter\def\csname res:passdiffhi:class:D:C\endcsname{-4.3}
\expandafter\def\csname res:passp:class:D:C\endcsname{0.002}
\expandafter\def\csname res:costratio:class:D:A\endcsname{1.28}
\expandafter\def\csname res:costratiolo:class:D:A\endcsname{1.21}
\expandafter\def\csname res:costratiohi:class:D:A\endcsname{1.35}
\expandafter\def\csname res:costp:class:D:A\endcsname{\ensuremath{<10^{-4}}}
\expandafter\def\csname res:wallratio:class:D:A\endcsname{2.50}
\expandafter\def\csname res:wallratiolo:class:D:A\endcsname{2.35}
\expandafter\def\csname res:wallratiohi:class:D:A\endcsname{2.65}
\expandafter\def\csname res:wallp:class:D:A\endcsname{\ensuremath{<10^{-4}}}
\expandafter\def\csname res:passdiff:class:D:A\endcsname{-12.9}
\expandafter\def\csname res:passdifflo:class:D:A\endcsname{-19.6}
\expandafter\def\csname res:passdiffhi:class:D:A\endcsname{-6.7}
\expandafter\def\csname res:passp:class:D:A\endcsname{\ensuremath{<10^{-4}}}
\expandafter\def\csname res:costratio:class:Aprime:A\endcsname{0.04}
\expandafter\def\csname res:costratiolo:class:Aprime:A\endcsname{0.04}
\expandafter\def\csname res:costratiohi:class:Aprime:A\endcsname{0.04}
\expandafter\def\csname res:costp:class:Aprime:A\endcsname{\ensuremath{<10^{-4}}}
\expandafter\def\csname res:wallratio:class:Aprime:A\endcsname{0.81}
\expandafter\def\csname res:wallratiolo:class:Aprime:A\endcsname{0.74}
\expandafter\def\csname res:wallratiohi:class:Aprime:A\endcsname{0.91}
\expandafter\def\csname res:wallp:class:Aprime:A\endcsname{\ensuremath{<10^{-4}}}
\expandafter\def\csname res:passdiff:class:Aprime:A\endcsname{-37.6}
\expandafter\def\csname res:passdifflo:class:Aprime:A\endcsname{-47.5}
\expandafter\def\csname res:passdiffhi:class:Aprime:A\endcsname{-28.2}
\expandafter\def\csname res:passp:class:Aprime:A\endcsname{\ensuremath{<10^{-4}}}
\expandafter\def\csname res:costratio:class:R:A\endcsname{0.19}
\expandafter\def\csname res:costratiolo:class:R:A\endcsname{0.12}
\expandafter\def\csname res:costratiohi:class:R:A\endcsname{0.27}
\expandafter\def\csname res:costp:class:R:A\endcsname{\ensuremath{<10^{-4}}}
\expandafter\def\csname res:wallratio:class:R:A\endcsname{0.99}
\expandafter\def\csname res:wallratiolo:class:R:A\endcsname{0.91}
\expandafter\def\csname res:wallratiohi:class:R:A\endcsname{1.07}
\expandafter\def\csname res:wallp:class:R:A\endcsname{0.24}
\expandafter\def\csname res:passdiff:class:R:A\endcsname{-34.9}
\expandafter\def\csname res:passdifflo:class:R:A\endcsname{-45.1}
\expandafter\def\csname res:passdiffhi:class:R:A\endcsname{-25.1}
\expandafter\def\csname res:passp:class:R:A\endcsname{\ensuremath{<10^{-4}}}
\expandafter\def\csname res:costratio:class:D:Aprime\endcsname{31.2}
\expandafter\def\csname res:costratiolo:class:D:Aprime\endcsname{28.6}
\expandafter\def\csname res:costratiohi:class:D:Aprime\endcsname{33.7}
\expandafter\def\csname res:costp:class:D:Aprime\endcsname{\ensuremath{<10^{-4}}}
\expandafter\def\csname res:wallratio:class:D:Aprime\endcsname{3.07}
\expandafter\def\csname res:wallratiolo:class:D:Aprime\endcsname{2.82}
\expandafter\def\csname res:wallratiohi:class:D:Aprime\endcsname{3.32}
\expandafter\def\csname res:wallp:class:D:Aprime\endcsname{\ensuremath{<10^{-4}}}
\expandafter\def\csname res:passdiff:class:D:Aprime\endcsname{+24.7}
\expandafter\def\csname res:passdifflo:class:D:Aprime\endcsname{+16.5}
\expandafter\def\csname res:passdiffhi:class:D:Aprime\endcsname{+33.3}
\expandafter\def\csname res:passp:class:D:Aprime\endcsname{\ensuremath{<10^{-4}}}
\expandafter\def\csname res:costratio:class:R:D\endcsname{0.15}
\expandafter\def\csname res:costratiolo:class:R:D\endcsname{0.10}
\expandafter\def\csname res:costratiohi:class:R:D\endcsname{0.21}
\expandafter\def\csname res:costp:class:R:D\endcsname{\ensuremath{<10^{-4}}}
\expandafter\def\csname res:wallratio:class:R:D\endcsname{0.39}
\expandafter\def\csname res:wallratiolo:class:R:D\endcsname{0.37}
\expandafter\def\csname res:wallratiohi:class:R:D\endcsname{0.42}
\expandafter\def\csname res:wallp:class:R:D\endcsname{\ensuremath{<10^{-4}}}
\expandafter\def\csname res:passdiff:class:R:D\endcsname{-22.0}
\expandafter\def\csname res:passdifflo:class:R:D\endcsname{-31.4}
\expandafter\def\csname res:passdiffhi:class:R:D\endcsname{-12.9}
\expandafter\def\csname res:passp:class:R:D\endcsname{\ensuremath{<10^{-4}}}
\expandafter\def\csname res:table:pairs:class\endcsname{B vs A & 3.03 [2.90, 3.16] & \ensuremath{<10^{-4}} & 2.96 [2.83, 3.10] & \ensuremath{<10^{-4}} & -0.8 [-5.1, +3.9] & 0.33 \\
C vs A & 3.05 [2.93, 3.19] & \ensuremath{<10^{-4}} & 3.01 [2.88, 3.15] & \ensuremath{<10^{-4}} & -2.4 [-6.3, +1.2] & 0.28 \\
C vs B & 1.01 [0.99, 1.02] & 0.29 & 1.02 [1.00, 1.03] & 0.068 & -1.6 [-5.5, +2.4] & 0.38 \\
D vs C & 0.42 [0.40, 0.44] & \ensuremath{<10^{-4}} & 0.83 [0.79, 0.87] & \ensuremath{<10^{-4}} & -10.6 [-16.9, -4.3] & 0.002 \\
D vs A & 1.28 [1.21, 1.35] & \ensuremath{<10^{-4}} & 2.50 [2.35, 2.65] & \ensuremath{<10^{-4}} & -12.9 [-19.6, -6.7] & \ensuremath{<10^{-4}} \\
A$'$ vs A & 0.04 [0.04, 0.04] & \ensuremath{<10^{-4}} & 0.81 [0.74, 0.91] & \ensuremath{<10^{-4}} & -37.6 [-47.5, -28.2] & \ensuremath{<10^{-4}} \\
R vs A & 0.19 [0.12, 0.27] & \ensuremath{<10^{-4}} & 0.99 [0.91, 1.07] & 0.24 & -34.9 [-45.1, -25.1] & \ensuremath{<10^{-4}} \\
D vs A$'$ & 31.2 [28.6, 33.7] & \ensuremath{<10^{-4}} & 3.07 [2.82, 3.32] & \ensuremath{<10^{-4}} & +24.7 [+16.5, +33.3] & \ensuremath{<10^{-4}} \\
R vs D & 0.15 [0.10, 0.21] & \ensuremath{<10^{-4}} & 0.39 [0.37, 0.42] & \ensuremath{<10^{-4}} & -22.0 [-31.4, -12.9] & \ensuremath{<10^{-4}} \\}
\expandafter\def\csname res:costratio:all:B:A\endcsname{3.20}
\expandafter\def\csname res:costratiolo:all:B:A\endcsname{3.07}
\expandafter\def\csname res:costratiohi:all:B:A\endcsname{3.35}
\expandafter\def\csname res:costp:all:B:A\endcsname{\ensuremath{<10^{-4}}}
\expandafter\def\csname res:wallratio:all:B:A\endcsname{3.20}
\expandafter\def\csname res:wallratiolo:all:B:A\endcsname{3.06}
\expandafter\def\csname res:wallratiohi:all:B:A\endcsname{3.35}
\expandafter\def\csname res:wallp:all:B:A\endcsname{\ensuremath{<10^{-4}}}
\expandafter\def\csname res:passdiff:all:B:A\endcsname{-0.5}
\expandafter\def\csname res:passdifflo:all:B:A\endcsname{-3.2}
\expandafter\def\csname res:passdiffhi:all:B:A\endcsname{+2.5}
\expandafter\def\csname res:passp:all:B:A\endcsname{0.31}
\expandafter\def\csname res:costratio:all:C:A\endcsname{3.22}
\expandafter\def\csname res:costratiolo:all:C:A\endcsname{3.10}
\expandafter\def\csname res:costratiohi:all:C:A\endcsname{3.37}
\expandafter\def\csname res:costp:all:C:A\endcsname{\ensuremath{<10^{-4}}}
\expandafter\def\csname res:wallratio:all:C:A\endcsname{3.24}
\expandafter\def\csname res:wallratiolo:all:C:A\endcsname{3.10}
\expandafter\def\csname res:wallratiohi:all:C:A\endcsname{3.39}
\expandafter\def\csname res:wallp:all:C:A\endcsname{\ensuremath{<10^{-4}}}
\expandafter\def\csname res:passdiff:all:C:A\endcsname{-1.5}
\expandafter\def\csname res:passdifflo:all:C:A\endcsname{-4.0}
\expandafter\def\csname res:passdiffhi:all:C:A\endcsname{+1.0}
\expandafter\def\csname res:passp:all:C:A\endcsname{0.28}
\expandafter\def\csname res:costratio:all:C:B\endcsname{1.01}
\expandafter\def\csname res:costratiolo:all:C:B\endcsname{0.99}
\expandafter\def\csname res:costratiohi:all:C:B\endcsname{1.02}
\expandafter\def\csname res:costp:all:C:B\endcsname{0.36}
\expandafter\def\csname res:wallratio:all:C:B\endcsname{1.01}
\expandafter\def\csname res:wallratiolo:all:C:B\endcsname{1.00}
\expandafter\def\csname res:wallratiohi:all:C:B\endcsname{1.03}
\expandafter\def\csname res:wallp:all:C:B\endcsname{0.15}
\expandafter\def\csname res:passdiff:all:C:B\endcsname{-1.0}
\expandafter\def\csname res:passdifflo:all:C:B\endcsname{-3.5}
\expandafter\def\csname res:passdiffhi:all:C:B\endcsname{+1.5}
\expandafter\def\csname res:passp:all:C:B\endcsname{0.38}
\expandafter\def\csname res:costratio:all:D:C\endcsname{0.39}
\expandafter\def\csname res:costratiolo:all:D:C\endcsname{0.38}
\expandafter\def\csname res:costratiohi:all:D:C\endcsname{0.41}
\expandafter\def\csname res:costp:all:D:C\endcsname{\ensuremath{<10^{-4}}}
\expandafter\def\csname res:wallratio:all:D:C\endcsname{0.80}
\expandafter\def\csname res:wallratiolo:all:D:C\endcsname{0.77}
\expandafter\def\csname res:wallratiohi:all:D:C\endcsname{0.84}
\expandafter\def\csname res:wallp:all:D:C\endcsname{\ensuremath{<10^{-4}}}
\expandafter\def\csname res:passdiff:all:D:C\endcsname{-6.7}
\expandafter\def\csname res:passdifflo:all:D:C\endcsname{-11.1}
\expandafter\def\csname res:passdiffhi:all:D:C\endcsname{-3.0}
\expandafter\def\csname res:passp:all:D:C\endcsname{0.002}
\expandafter\def\csname res:costratio:all:D:A\endcsname{1.27}
\expandafter\def\csname res:costratiolo:all:D:A\endcsname{1.21}
\expandafter\def\csname res:costratiohi:all:D:A\endcsname{1.33}
\expandafter\def\csname res:costp:all:D:A\endcsname{\ensuremath{<10^{-4}}}
\expandafter\def\csname res:wallratio:all:D:A\endcsname{2.60}
\expandafter\def\csname res:wallratiolo:all:D:A\endcsname{2.46}
\expandafter\def\csname res:wallratiohi:all:D:A\endcsname{2.74}
\expandafter\def\csname res:wallp:all:D:A\endcsname{\ensuremath{<10^{-4}}}
\expandafter\def\csname res:passdiff:all:D:A\endcsname{-8.1}
\expandafter\def\csname res:passdifflo:all:D:A\endcsname{-12.3}
\expandafter\def\csname res:passdiffhi:all:D:A\endcsname{-4.2}
\expandafter\def\csname res:passp:all:D:A\endcsname{\ensuremath{<10^{-4}}}
\expandafter\def\csname res:costratio:all:Aprime:A\endcsname{0.04}
\expandafter\def\csname res:costratiolo:all:Aprime:A\endcsname{0.04}
\expandafter\def\csname res:costratiohi:all:Aprime:A\endcsname{0.04}
\expandafter\def\csname res:costp:all:Aprime:A\endcsname{\ensuremath{<10^{-4}}}
\expandafter\def\csname res:wallratio:all:Aprime:A\endcsname{0.79}
\expandafter\def\csname res:wallratiolo:all:Aprime:A\endcsname{0.73}
\expandafter\def\csname res:wallratiohi:all:Aprime:A\endcsname{0.87}
\expandafter\def\csname res:wallp:all:Aprime:A\endcsname{\ensuremath{<10^{-4}}}
\expandafter\def\csname res:passdiff:all:Aprime:A\endcsname{-24.2}
\expandafter\def\csname res:passdifflo:all:Aprime:A\endcsname{-31.4}
\expandafter\def\csname res:passdiffhi:all:Aprime:A\endcsname{-17.5}
\expandafter\def\csname res:passp:all:Aprime:A\endcsname{\ensuremath{<10^{-4}}}
\expandafter\def\csname res:costratio:all:R:A\endcsname{0.18}
\expandafter\def\csname res:costratiolo:all:R:A\endcsname{0.12}
\expandafter\def\csname res:costratiohi:all:R:A\endcsname{0.24}
\expandafter\def\csname res:costp:all:R:A\endcsname{\ensuremath{<10^{-4}}}
\expandafter\def\csname res:wallratio:all:R:A\endcsname{1.00}
\expandafter\def\csname res:wallratiolo:all:R:A\endcsname{0.93}
\expandafter\def\csname res:wallratiohi:all:R:A\endcsname{1.07}
\expandafter\def\csname res:wallp:all:R:A\endcsname{0.29}
\expandafter\def\csname res:passdiff:all:R:A\endcsname{-22.5}
\expandafter\def\csname res:passdifflo:all:R:A\endcsname{-29.4}
\expandafter\def\csname res:passdiffhi:all:R:A\endcsname{-15.8}
\expandafter\def\csname res:passp:all:R:A\endcsname{\ensuremath{<10^{-4}}}
\expandafter\def\csname res:costratio:all:D:Aprime\endcsname{32.0}
\expandafter\def\csname res:costratiolo:all:D:Aprime\endcsname{29.5}
\expandafter\def\csname res:costratiohi:all:D:Aprime\endcsname{34.3}
\expandafter\def\csname res:costp:all:D:Aprime\endcsname{\ensuremath{<10^{-4}}}
\expandafter\def\csname res:wallratio:all:D:Aprime\endcsname{3.29}
\expandafter\def\csname res:wallratiolo:all:D:Aprime\endcsname{3.04}
\expandafter\def\csname res:wallratiohi:all:D:Aprime\endcsname{3.56}
\expandafter\def\csname res:wallp:all:D:Aprime\endcsname{\ensuremath{<10^{-4}}}
\expandafter\def\csname res:passdiff:all:D:Aprime\endcsname{+16.0}
\expandafter\def\csname res:passdifflo:all:D:Aprime\endcsname{+10.6}
\expandafter\def\csname res:passdiffhi:all:D:Aprime\endcsname{+22.0}
\expandafter\def\csname res:passp:all:D:Aprime\endcsname{\ensuremath{<10^{-4}}}
\expandafter\def\csname res:costratio:all:R:D\endcsname{0.14}
\expandafter\def\csname res:costratiolo:all:R:D\endcsname{0.09}
\expandafter\def\csname res:costratiohi:all:R:D\endcsname{0.19}
\expandafter\def\csname res:costp:all:R:D\endcsname{\ensuremath{<10^{-4}}}
\expandafter\def\csname res:wallratio:all:R:D\endcsname{0.38}
\expandafter\def\csname res:wallratiolo:all:R:D\endcsname{0.36}
\expandafter\def\csname res:wallratiohi:all:R:D\endcsname{0.40}
\expandafter\def\csname res:wallp:all:R:D\endcsname{\ensuremath{<10^{-4}}}
\expandafter\def\csname res:passdiff:all:R:D\endcsname{-14.3}
\expandafter\def\csname res:passdifflo:all:R:D\endcsname{-20.7}
\expandafter\def\csname res:passdiffhi:all:R:D\endcsname{-8.4}
\expandafter\def\csname res:passp:all:R:D\endcsname{\ensuremath{<10^{-4}}}
\expandafter\def\csname res:table:pairs:all\endcsname{B vs A & 3.20 [3.07, 3.35] & \ensuremath{<10^{-4}} & 3.20 [3.06, 3.35] & \ensuremath{<10^{-4}} & -0.5 [-3.2, +2.5] & 0.31 \\
C vs A & 3.22 [3.10, 3.37] & \ensuremath{<10^{-4}} & 3.24 [3.10, 3.39] & \ensuremath{<10^{-4}} & -1.5 [-4.0, +1.0] & 0.28 \\
C vs B & 1.01 [0.99, 1.02] & 0.36 & 1.01 [1.00, 1.03] & 0.15 & -1.0 [-3.5, +1.5] & 0.38 \\
D vs C & 0.39 [0.38, 0.41] & \ensuremath{<10^{-4}} & 0.80 [0.77, 0.84] & \ensuremath{<10^{-4}} & -6.7 [-11.1, -3.0] & 0.002 \\
D vs A & 1.27 [1.21, 1.33] & \ensuremath{<10^{-4}} & 2.60 [2.46, 2.74] & \ensuremath{<10^{-4}} & -8.1 [-12.3, -4.2] & \ensuremath{<10^{-4}} \\
A$'$ vs A & 0.04 [0.04, 0.04] & \ensuremath{<10^{-4}} & 0.79 [0.73, 0.87] & \ensuremath{<10^{-4}} & -24.2 [-31.4, -17.5] & \ensuremath{<10^{-4}} \\
R vs A & 0.18 [0.12, 0.24] & \ensuremath{<10^{-4}} & 1.00 [0.93, 1.07] & 0.29 & -22.5 [-29.4, -15.8] & \ensuremath{<10^{-4}} \\
D vs A$'$ & 32.0 [29.5, 34.3] & \ensuremath{<10^{-4}} & 3.29 [3.04, 3.56] & \ensuremath{<10^{-4}} & +16.0 [+10.6, +22.0] & \ensuremath{<10^{-4}} \\
R vs D & 0.14 [0.09, 0.19] & \ensuremath{<10^{-4}} & 0.38 [0.36, 0.40] & \ensuremath{<10^{-4}} & -14.3 [-20.7, -8.4] & \ensuremath{<10^{-4}} \\}
\expandafter\def\csname res:tau:median\endcsname{7.2}
\expandafter\def\csname res:tau:mean\endcsname{7.5}
\expandafter\def\csname res:tau:lo\endcsname{7.4}
\expandafter\def\csname res:tau:hi\endcsname{7.6}
\expandafter\def\csname res:tau:p95\endcsname{12.5}
\expandafter\def\csname res:tau:max\endcsname{24}
\expandafter\def\csname res:tau:turns\endcsname{2430}
\expandafter\def\csname res:tau:pipeline\endcsname{20.5}
\expandafter\def\csname res:tau:share\endcsname{0.12}
\expandafter\def\csname res:envelope:kb\endcsname{19.5}
\expandafter\def\csname res:tau:shareC\endcsname{0.11}
\expandafter\def\csname res:cb:diff\endcsname{+0.29}
\expandafter\def\csname res:cb:lo\endcsname{-0.04}
\expandafter\def\csname res:cb:hi\endcsname{+0.62}
\expandafter\def\csname res:cb:p\endcsname{0.15}
\expandafter\def\csname res:model:alphaout:function\endcsname{4.18}
\expandafter\def\csname res:model:alphain:function\endcsname{5.56}
\expandafter\def\csname res:model:rho:function\endcsname{0.17}
\expandafter\def\csname res:model:fout:function\endcsname{0.83}
\expandafter\def\csname res:model:fin:function\endcsname{0.61}
\expandafter\def\csname res:model:fstar:function\endcsname{0.78}
\expandafter\def\csname res:model:predout:function\endcsname{0.80}
\expandafter\def\csname res:model:predfixed:function\endcsname{1.26}
\expandafter\def\csname res:model:observed:function\endcsname{1.17}
\expandafter\def\csname res:model:alphaoutD:function\endcsname{5.29}
\expandafter\def\csname res:model:alphainD:function\endcsname{4.82}
\expandafter\def\csname res:model:volume:function\endcsname{1.18}
\expandafter\def\csname res:model:price:function\endcsname{0.23}
\expandafter\def\csname res:model:codershare:function\endcsname{90.7}
\expandafter\def\csname res:model:forecasterr:function\endcsname{+7}
\expandafter\def\csname res:model:alphaout:class\endcsname{2.78}
\expandafter\def\csname res:model:alphain:class\endcsname{4.77}
\expandafter\def\csname res:model:rho:class\endcsname{0.14}
\expandafter\def\csname res:model:fout:class\endcsname{0.73}
\expandafter\def\csname res:model:fin:class\endcsname{0.60}
\expandafter\def\csname res:model:fstar:class\endcsname{0.66}
\expandafter\def\csname res:model:predout:class\endcsname{0.80}
\expandafter\def\csname res:model:predfixed:class\endcsname{0.99}
\expandafter\def\csname res:model:observed:class\endcsname{1.28}
\expandafter\def\csname res:model:alphaoutD:class\endcsname{3.93}
\expandafter\def\csname res:model:alphainD:class\endcsname{5.19}
\expandafter\def\csname res:model:volume:class\endcsname{1.35}
\expandafter\def\csname res:model:price:class\endcsname{0.31}
\expandafter\def\csname res:model:codershare:class\endcsname{94.2}
\expandafter\def\csname res:model:forecasterr:class\endcsname{-23}
\expandafter\def\csname res:model:alphaout:all\endcsname{2.95}
\expandafter\def\csname res:model:alphain:all\endcsname{4.89}
\expandafter\def\csname res:model:rho:all\endcsname{0.14}
\expandafter\def\csname res:model:fout:all\endcsname{0.74}
\expandafter\def\csname res:model:fin:all\endcsname{0.60}
\expandafter\def\csname res:model:fstar:all\endcsname{0.68}
\expandafter\def\csname res:model:predout:all\endcsname{0.81}
\expandafter\def\csname res:model:predfixed:all\endcsname{1.02}
\expandafter\def\csname res:model:observed:all\endcsname{1.27}
\expandafter\def\csname res:model:alphaoutD:all\endcsname{4.10}
\expandafter\def\csname res:model:alphainD:all\endcsname{5.13}
\expandafter\def\csname res:model:volume:all\endcsname{1.32}
\expandafter\def\csname res:model:price:all\endcsname{0.30}
\expandafter\def\csname res:model:codershare:all\endcsname{92.9}
\expandafter\def\csname res:model:forecasterr:all\endcsname{-19}
\expandafter\def\csname res:router:small:function\endcsname{98.0}
\expandafter\def\csname res:router:overhead:function\endcsname{25.2}
\expandafter\def\csname res:oracle:pass:function\endcsname{100.0}
\expandafter\def\csname res:oracle:cps:function\endcsname{0.00034}
\expandafter\def\csname res:oracle:cheap:function\endcsname{98.0}
\expandafter\def\csname res:router:smallfail:function\endcsname{1.4}
\expandafter\def\csname res:router:largewaste:function\endcsname{100.0}
\expandafter\def\csname res:router:small:class\endcsname{86.7}
\expandafter\def\csname res:router:overhead:class\endcsname{9.2}
\expandafter\def\csname res:oracle:pass:class\endcsname{91.0}
\expandafter\def\csname res:oracle:cps:class\endcsname{0.0159}
\expandafter\def\csname res:oracle:cheap:class\endcsname{54.1}
\expandafter\def\csname res:router:smallfail:class\endcsname{47.1}
\expandafter\def\csname res:router:largewaste:class\endcsname{44.1}
\expandafter\def\csname res:router:small:all\endcsname{90.9}
\expandafter\def\csname res:router:overhead:all\endcsname{15.1}
\expandafter\def\csname res:oracle:pass:all\endcsname{94.3}
\expandafter\def\csname res:oracle:cps:all\endcsname{0.0098}
\expandafter\def\csname res:oracle:cheap:all\endcsname{70.4}
\expandafter\def\csname res:router:smallfail:all\endcsname{28.8}
\expandafter\def\csname res:router:largewaste:all\endcsname{48.6}
\expandafter\def\csname res:tps:strong\endcsname{115}
\expandafter\def\csname res:tps:cheap\endcsname{244}
\expandafter\def\csname res:corr:all\endcsname{0.87}
\expandafter\def\csname res:corr:withinlo\endcsname{0.91}
\expandafter\def\csname res:corr:withinhi\endcsname{0.98}
\expandafter\def\csname res:fail:compile:A\endcsname{1}
\expandafter\def\csname res:fail:timeout:A\endcsname{0}
\expandafter\def\csname res:fail:truncated:A\endcsname{0}
\expandafter\def\csname res:fail:refusal:A\endcsname{0}
\expandafter\def\csname res:fail:tests:A\endcsname{24}
\expandafter\def\csname res:fail:compile:Aprime\endcsname{2}
\expandafter\def\csname res:fail:timeout:Aprime\endcsname{0}
\expandafter\def\csname res:fail:truncated:Aprime\endcsname{0}
\expandafter\def\csname res:fail:refusal:Aprime\endcsname{0}
\expandafter\def\csname res:fail:tests:Aprime\endcsname{121}
\expandafter\def\csname res:fail:compile:R\endcsname{1}
\expandafter\def\csname res:fail:timeout:R\endcsname{0}
\expandafter\def\csname res:fail:truncated:R\endcsname{0}
\expandafter\def\csname res:fail:refusal:R\endcsname{0}
\expandafter\def\csname res:fail:tests:R\endcsname{115}
\expandafter\def\csname res:fail:compile:B\endcsname{0}
\expandafter\def\csname res:fail:timeout:B\endcsname{0}
\expandafter\def\csname res:fail:truncated:B\endcsname{0}
\expandafter\def\csname res:fail:refusal:B\endcsname{1}
\expandafter\def\csname res:fail:tests:B\endcsname{27}
\expandafter\def\csname res:fail:compile:C\endcsname{1}
\expandafter\def\csname res:fail:timeout:C\endcsname{0}
\expandafter\def\csname res:fail:truncated:C\endcsname{0}
\expandafter\def\csname res:fail:refusal:C\endcsname{0}
\expandafter\def\csname res:fail:tests:C\endcsname{30}
\expandafter\def\csname res:fail:compile:D\endcsname{0}
\expandafter\def\csname res:fail:timeout:D\endcsname{0}
\expandafter\def\csname res:fail:truncated:D\endcsname{0}
\expandafter\def\csname res:fail:refusal:D\endcsname{0}
\expandafter\def\csname res:fail:tests:D\endcsname{58}
\expandafter\def\csname res:H1\endcsname{supported}
\expandafter\def\csname res:H2\endcsname{not supported}
\expandafter\def\csname res:H3\endcsname{supported}
\expandafter\def\csname res:H4\endcsname{not supported}

\newcommand{\ressum}[1]{\ifcsname res:fail:#1:A\endcsname\the\numexpr\csname res:fail:#1:A\endcsname+\csname res:fail:#1:Aprime\endcsname+\csname res:fail:#1:R\endcsname+\csname res:fail:#1:B\endcsname+\csname res:fail:#1:C\endcsname+\csname res:fail:#1:D\endcsname\relax\else\textbf{\textcolor{red}{??}}\fi}

\title{The Coordination Tax and the Heterogeneity Dividend: Break-Even Conditions for Multi-Agent LLM Pipelines}

\author{
	\href{https://orcid.org/0009-0001-3478-198X}{\includegraphics[scale=0.06]{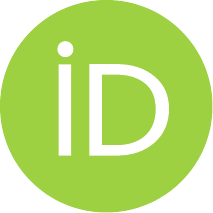}\hspace{1mm}Andreas Zettl}\\
	Member, VoiceInteroperability.ai\\
	Linux Foundation AI \& Data\\
	Berlin, Germany\\
	\texttt{info@azettl.net}
	\and
	\href{https://orcid.org/0009-0008-7513-1255}{\includegraphics[scale=0.06]{orcid.pdf}\hspace{1mm}Diego Gosmar}\\
	Member, VoiceInteroperability.ai\\
	Linux Foundation AI \& Data\\
	Turin, Italy\\
	\texttt{diego.gosmar@ieee.org}
}

\begin{document}

	\maketitle
\csname res:banner\endcsname

\begin{abstract}
Multi-agent LLM pipelines built on interoperability protocols such as the Open Floor Protocol (OFP) or A2A are usually justified on capability grounds, with their runtime cost treated as an overhead of unknown size. Comparing such a pipeline with one model call is underspecified, because a pipeline can give each role a different model. We give a closed-form model that sets the coordination tax -- decomposition overhead plus per-handoff protocol cost -- against the heterogeneity dividend from routing work to a cheaper model, with a break-even threshold and a feasibility bound, $\alpha < 1/r$, and measure every term, with OFP as the protocol under test, on \res{tasks} HumanEval and ClassEval tasks in \res{runs} runs of six architectures with Claude Opus~5.5 and Haiku~5.5. The protocol adds \res{tau:median}\,ms per handoff, \res{tau:shareC}\% of pipeline wall time. Decomposition does not pay on these tasks: a planner--coder--reviewer pipeline costs \res{costratio:all:C:A}$\times$ as much as a single strong call, takes \res{wallratio:all:C:A}$\times$ as long, and solves no more tasks. Moving planner and reviewer to the cheap model cuts the pipeline's cost to \res{costratio:all:D:C}$\times$ but changes pass@1 by \res{passdiff:all:D:C} points, leaving it dearer, slower and less accurate than the single call, and the break-even forecast misses by \res{model:forecasterr:all}\% because the cheap roles inflate token volume. The saving lies in choosing a model per task instead: an oracle choice between the two single calls solves \res{oracle:pass:all}\% of runs at half the strong call's cost per solved task.

\end{abstract}

\section{Introduction}

Interoperability protocols for autonomous agents are arriving faster than the evidence for them. The Open Floor Protocol (OFP) \cite{ofp2025} standardises a conversational floor over which heterogeneous agents exchange JSON envelopes; A2A \cite{a2a2025} standardises peer-to-peer agent task delegation; the Model Context Protocol \cite{mcp2024} standardises the agent-to-tool boundary. Each is argued for on the same grounds: agents built on different stacks can discover one another, compose across organisational boundaries, and be replaced independently. Each is deployed on the tacit assumption that the runtime price of that flexibility is small enough not to matter.

The assumption is rarely tested. Practitioners ask a blunter question instead -- is a multi-agent pipeline simply slower and more expensive than one call to the model? -- and the answer usually offered is a benchmark number for one task, one model and one architecture.

The question matters most for voice agents. OFP grew out of the effort to let conversational and voice assistants from different vendors work together \cite{ovoninter,gosmar2024interop,gosmar2024nl4ai}, and a voice agent is judged turn by turn: whatever time the agents spend coordinating behind the scenes delays the spoken reply the user is waiting for. For such agents latency counts as much as spend, and any overhead that a protocol or a decomposition adds is paid inside the conversational turn.

That comparison conflates three mechanisms that move cost in different directions and at different magnitudes. \emph{Decomposition} splits one generation into several, re-establishing context at every stage and inflating total token volume. The \emph{protocol layer} adds serialisation, envelope construction, transport and floor management at every handoff. \emph{Model heterogeneity} allows each role to be served by a different model, so that mechanical roles can be pushed onto cheaper and faster models while only the roles that need frontier capability pay frontier prices. Only the third mechanism can be negative, and it is available exclusively to the decomposed architecture. A benchmark that holds the model fixed across pipeline stages therefore measures the cost of decomposition and the protocol tax while deliberately discarding the only source of saving -- and then reports the sum as the cost of the protocol.

This paper reframes the question as an economic one. We ask under what conditions the heterogeneity dividend exceeds the coordination tax, derive those conditions in closed form, and measure every parameter in them. The target is a statement of the form: for a decomposition overhead of $\alpha$, a cheap/strong price ratio of $r$, and a routable work fraction $f$, protocol-mediated decomposition is cheaper than a single strong call if and only if $f$ exceeds an explicit threshold -- and is never cheaper, at any $f$, once $\alpha$ exceeds $1/r$. That statement is protocol-agnostic and survives the next protocol, which a raw benchmark number does not. It is also falsifiable, and we test it. Our contributions are:

\begin{itemize}
\item a closed-form model of the coordination tax and the heterogeneity dividend, with a cost break-even threshold, a feasibility bound, a latency condition and a misrouting term, and a reporting rule that splits any heterogeneous-versus-uniform cost ratio into a volume factor and a price factor (Section~\ref{sec:model});
\item a six-condition design that separates decomposition, protocol and model assignment, and includes the two baselines an economic claim has to beat: one call on the cheap model, and a router (Section~\ref{sec:design});
\item a direct measurement of an interoperability protocol's per-handoff runtime overhead, timed inside the convener rather than inferred from end-to-end differences (Section~\ref{sec:results});
\item results on \res{tasks} tasks in two difficulty tiers, analysed with the task as the unit and against hypotheses fixed before data collection (Sections~\ref{sec:design}--\ref{sec:discussion}).
\end{itemize}

Section~\ref{sec:related} positions the work, Section~\ref{sec:model} develops the model, Section~\ref{sec:design} specifies the study, and Sections~\ref{sec:results} and~\ref{sec:discussion} report and discuss the results. Sections~\ref{sec:threats}--\ref{sec:integrity} cover validity, future work and reproducibility.

\section{Background and Related Work}\label{sec:related}

\subsection{Agent Interoperability Protocols}

The Open Floor Protocol, developed within the Open Voice Interoperability Initiative of the Linux Foundation AI \& Data Foundation, standardises the Conversation Envelope that every agent sends and receives, Dialog Events representing what was said, Assistant Manifests describing what an agent can do, and a set of floor management events -- \texttt{requestFloor}, \texttt{grantFloor}, \texttt{revokeFloor}, \texttt{yieldFloor} -- by which agents built by different teams on different stacks take turns in a shared conversation \cite{ofp2025,ovoninter,ovonspecs2024}. Its organising metaphor is multi-party conversation: agents gather around a shared floor, and floor management -- who holds it, who is invited, who is delegated to -- is part of the protocol rather than of the application. This is a deliberate departure from point-to-point delegation models, and it is the feature that makes mid-task reassignment expressible at the protocol level rather than at the orchestrator level; the conception of conversational agency underlying that choice is developed in \cite{hyperconv}. The Assistant Manifest gives discovery a declarative form, so that adding a capability is in principle a manifest change rather than a redeployment. Extensions to the floor model for managing multiparty conversations among heterogeneous assistants are developed in \cite{gosmar2024interop,gosmar2024nl4ai}, which sets out the delegation and convening semantics the present pipelines rely on, building on the universal open APIs for agentic multimodal communication proposed in \cite{gosmar2024openapis}.

The intellectual ancestry runs through agent communication languages such as FIPA ACL \cite{fipa2002}, which likewise separated message structure and interaction protocol from agent internals. The modern reprise differs in that the agents are stochastic, the cost of a message is dominated by inference rather than transport, and the price of a message varies by two orders of magnitude depending on which model is behind it. That last property is what makes the economics non-trivial, and it has no analogue in the classical literature.

\subsection{Cost-Aware Serving: Cascades and Routers}

A parallel line of work treats model choice itself as the optimisation variable. FrugalGPT \cite{frugalgpt2023} cascades queries through progressively more expensive models with a learned stopping rule, reporting large cost reductions at matched quality. RouteLLM \cite{routellm2024} trains a router that dispatches each query to a strong or a weak model based on predicted difficulty. Both establish that heterogeneous serving is where the money is; neither involves decomposition, an inter-agent protocol, or more than one model per query. Semantic caching is a third lever on the same cost function, reducing generated volume rather than unit price \cite{liu2025semantic,gosmar2026aiai,gosmar2026hallucache}; it is orthogonal to the comparison studied here and we hold it out of all conditions, but it shares the property, central to Section~\ref{sec:pricevolume}, that savings in token volume and savings in token price must be accounted separately.

This literature supplies the competitor that agent-protocol benchmarks almost always omit. If the object is to save money by using a cheaper model where a cheaper model suffices, a single-call router achieves that with no decomposition overhead and no protocol tax whatsoever. Any claim that multi-agent routing is economically motivated must therefore be made against a router baseline, not against a single strong call. We treat its omission as the central methodological gap in existing comparisons.

\subsection{Multi-Agent LLM Frameworks and Their Evaluation}

Frameworks such as AutoGen \cite{autogen2023} and MetaGPT \cite{metagpt2023} popularised role-structured pipelines -- planner, implementer, reviewer -- and report quality gains on reasoning and code generation. Evaluation in this literature is predominantly quality-first, with cost and latency reported as secondary descriptive statistics when reported at all. Kapoor et al. \cite{kapoor2024} document the consequence: agent evaluations that do not control for cost systematically favour architectures that simply spend more inference, and accuracy-only leaderboards reward this. Their prescription -- report the accuracy-cost frontier rather than a point -- is the one we adopt, extended to latency and to the protocol layer specifically.

Work on staged agent pipelines with explicit inter-agent measurement is closer to our setting. An OFP-mediated three-stage pipeline is instrumented in \cite{icaart26} so that an evaluator observes the full trace without sitting in the decision path, precisely in order to make the protocol boundary measurable rather than implicit; related work on sentinel agents in multi-agent systems \cite{gosmar2025sentinel} adopts the same separation between the agents doing the work and the agents observing it, and the Agent Status Protocol \cite{gosmar2026asp} proposes a lightweight specification through which agents report their state to such observers. The present study extends that instrumentation from a single pipeline to a controlled comparison of architectures and model assignments.

\subsection{The Gap}

What remains absent from the literature, to our knowledge, is a design that separates decomposition, protocol and model assignment as independent factors; that times the protocol's per-handoff overhead directly instead of inferring it from noisy end-to-end differences; that includes the cheap single call and the router against which an economic claim must be made; and that evaluates on many tasks with quality off its ceiling, reporting cost per solved task rather than per attempt. That is the gap this study addresses. A comparison across protocols, and the fault-recovery, fan-out and discovery properties that motivate interoperability protocols in the first place, remain future work (Section~\ref{sec:future}).

\section{An Analytical Model of the Coordination Tax}\label{sec:model}

\subsection{Notation}

Consider a pipeline of $k$ stages with $H$ handoffs, where stage $i$ runs model $m_i$, consumes $\mathit{in}_i$ input tokens and emits $o_i$ output tokens, reasoning tokens included: they are billed and generated as output although they never appear in the response. Let $p_{\mathrm{in}}(m)$ and $p_{\mathrm{out}}(m)$ denote input and output token prices, $\ell(m,\mathit{in})$ the prefill or time-to-first-token, and $s(m)$ the generation throughput in tokens per second. Write $\tau$ for the per-handoff protocol tax. Latency and cost are then

\begin{align}
T_k &\;\approx\; \sum_{i=1}^{k} \left[ \ell(m_i, \mathit{in}_i) + \frac{o_i}{s(m_i)} \right] \;+\; H\,\tau , \label{eq:latency}\\
C_k &\;\approx\; \sum_{i=1}^{k} \left[ p_{\mathrm{in}}(m_i)\,\mathit{in}_i + p_{\mathrm{out}}(m_i)\,o_i \right] . \label{eq:cost}
\end{align}

The baseline is a single call on the strong model $m_s$ with $I_1$ input and $O_1$ output tokens, giving $T_1 = \ell_1 + O_1/s_s$ and $C_1 = p_{\mathrm{in}}^{s} I_1 + p_{\mathrm{out}}^{s} O_1$. Within one model, generation time is close to proportional to output length, which is the empirical warrant for treating \eqref{eq:latency} as tractable (Section~\ref{sec:latency} reports how close). $\tau$ is not assumed but measured: the convener times each handoff and subtracts the time the receiving agent spent inside its model call (Section~\ref{sec:instrumentation}).

\subsection{Three Derived Quantities}

Define the \emph{decomposition inflation factor} $\alpha = \left(\sum_i o_i\right)/O_1$, the total output token volume of the pipeline relative to the single call; the \emph{routable fraction} $f$, the share of pipeline output tokens generated by the cheap model; and the \emph{price ratio} $r = p_{\mathrm{out}}^{c}/p_{\mathrm{out}}^{s} \in (0,1)$. Each is a measurable property of a deployment rather than a modelling choice: $\alpha$ is set by how the task is cut and by which model fills each role (Section~\ref{sec:pricevolume}), $f$ by the routing policy, and $r$ by the provider's price list.

\subsection{The Cost Break-Even Condition}

Assume first that output tokens dominate cost, as they do at the price ratios of current frontier providers when prompts are short relative to completions. The pipeline is cheaper than the single strong call when

\begin{equation}
\alpha\left[(1-f) + f r\right] < 1
\qquad\Longleftrightarrow\qquad
f \;>\; \frac{\alpha - 1}{\alpha\,(1-r)} \;=:\; f^{\star}. \label{eq:breakeven}
\end{equation}

The two sides have a direct reading. The quantity $\alpha - 1$ is the \emph{decomposition tax}: the extra token volume the architecture generates merely by being decomposed. The quantity $\alpha f (1-r)$ is the \emph{heterogeneity dividend}: the saving obtained by serving a fraction $f$ of that inflated volume at a discount of $(1-r)$. Decomposition pays exactly when the dividend exceeds the tax, and \eqref{eq:breakeven} is that statement solved for $f$.

\subsection{A Feasibility Bound}

Because $f \le 1$ by construction, \eqref{eq:breakeven} admits a solution only if $f^{\star} < 1$, that is only if

\begin{equation}
\alpha\,r < 1 \qquad\Longleftrightarrow\qquad \alpha < 1/r . \label{eq:feasibility}
\end{equation}

This is the sharpest practical consequence of the model, and it is a hard ceiling rather than a tendency. Beyond it, no routing policy recovers the overhead -- not even pushing every token to the cheap model -- because the decomposed architecture is generating so much extra volume that the discount cannot catch up. With a cheap model at one-fifth the price of the strong one, decomposition may inflate token volume by up to a factor of five before routing becomes futile; with a cheap model at half price, the ceiling falls to a factor of two, which many role-structured pipelines exceed as a matter of course. At the price ratio of the models studied here, $r = 0.025$, the ceiling is a factor of forty, so feasibility is not the binding constraint and the threshold $f^{\star}$ is.

Table~\ref{tab:threshold} evaluates $f^{\star}$ over the parameter range of practical interest.

\begin{table}[!htbp]
\centering
\caption{Minimum routable fraction $f^{\star}$ required for a decomposed pipeline to undercut a single strong call, by decomposition inflation $\alpha$ and price ratio $r$. Dashes mark infeasible regions where $\alpha \ge 1/r$ and no routing policy suffices. The last column is the price ratio of the two models in this study.}\label{tab:threshold}
\begin{tabular}{lcccc}
\toprule
 & $r = 0.50$ & $r = 0.33$ & $r = 0.20$ & $r = 0.025$ \\
\midrule
$\alpha = 1.5$ & 0.67 & 0.50 & 0.42 & 0.34 \\
$\alpha = 2.0$ & 1.00 & 0.75 & 0.63 & 0.51 \\
$\alpha = 2.5$ & --- & 0.90 & 0.75 & 0.62 \\
$\alpha = 3.0$ & --- & 0.99 & 0.83 & 0.68 \\
$\alpha = 4.0$ & --- & --- & 0.94 & 0.77 \\
\midrule
feasibility ceiling $1/r$ & 2.00 & 3.03 & 5.00 & 40.0 \\
\bottomrule
\end{tabular}
\end{table}

\begin{figure}[!htbp]
\centering
\begin{tikzpicture}
\begin{axis}[
  width=10cm, height=6.2cm,
  xlabel={decomposition inflation $\alpha$},
  ylabel={required routable fraction $f^{\star}$},
  xmin=1, xmax=5.2, ymin=0, ymax=1.3,
  ytick={0,0.2,0.4,0.6,0.8,1.0},
  legend pos=south east,
  legend style={font=\scriptsize, draw=black!30},
  label style={font=\small},
  tick label style={font=\scriptsize},
  grid=both, grid style={black!12},
]
\addplot[domain=1:5.2, samples=200, thick, dashdotted] {(x-1)/(x*0.975)};
\addlegendentry{$r=0.025$ (this study)}
\addplot[domain=1:5, samples=200, thick] {(x-1)/(x*0.8)};
\addlegendentry{$r=0.20$}
\addplot[domain=1:3.03, samples=200, thick, dashed] {(x-1)/(x*0.67)};
\addlegendentry{$r=0.33$}
\addplot[domain=1:2, samples=200, thick, dotted] {(x-1)/(x*0.5)};
\addlegendentry{$r=0.50$}
\addplot[domain=1:5.2, samples=2, gray, thin] {1};
\node[font=\scriptsize, anchor=west, text=black!60] at (axis cs:1.15,1.16) {$f^{\star}=1$: no routing policy suffices above this line};
\end{axis}
\end{tikzpicture}
\caption{The break-even threshold $f^{\star} = (\alpha-1)/[\alpha(1-r)]$ as a function of decomposition inflation, for four price ratios. Each curve terminates where it meets $f^{\star}=1$, at $\alpha = 1/r$: to the right of that point the decomposed architecture cannot be made cheaper than a single strong call by any routing policy, however well the protocol is implemented. The curves are steepest near $\alpha = 1$, meaning that the first increments of decomposition overhead are the expensive ones.}\label{fig:breakeven}
\end{figure}
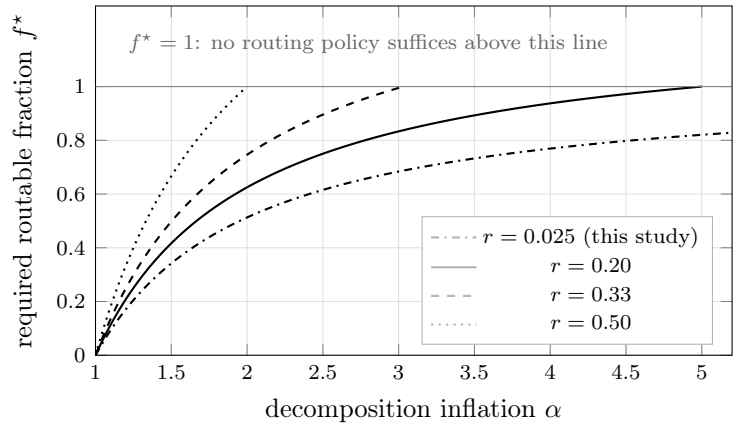

Table~\ref{tab:threshold} and Fig.~\ref{fig:breakeven} make the practical prediction explicit: at any realistic decomposition overhead, heterogeneous routing pays only when a clear majority of generated tokens can be pushed onto the cheap model. Pipelines that decompose aggressively while keeping a strong model in most roles are in the infeasible region regardless of how well the protocol is implemented, and no amount of protocol optimisation moves them out of it.

\subsection{Including Input Tokens}

Where prompts are long relative to completions -- retrieval-heavy or long-context settings -- the input side cannot be discarded. Let $\rho = p_{\mathrm{in}}^{s} I_1 / C_1$ be the input share of baseline cost, $\alpha_{\mathrm{in}}$ and $\alpha_{\mathrm{out}}$ the input and output inflation factors, $f_{\mathrm{in}}$ and $f_{\mathrm{out}}$ the corresponding routable fractions, and $r_{\mathrm{in}}, r_{\mathrm{out}}$ the corresponding price ratios. Then

\begin{equation}
\frac{C_k}{C_1} \;=\; \rho\,\alpha_{\mathrm{in}}\!\left[(1-f_{\mathrm{in}}) + f_{\mathrm{in}} r_{\mathrm{in}}\right] \;+\; (1-\rho)\,\alpha_{\mathrm{out}}\!\left[(1-f_{\mathrm{out}}) + f_{\mathrm{out}} r_{\mathrm{out}}\right].
\label{eq:full}
\end{equation}

Equation~\eqref{eq:breakeven} is the $\rho \to 0$ special case. The generalisation matters because $\alpha_{\mathrm{in}}$ is typically the larger of the two inflation factors: every stage after the first must be re-supplied with context that the single call held in one window, so input volume inflates roughly with pipeline depth even when output volume does not. Decomposition overhead is therefore worse, not better, in long-context regimes, and studies that report only output-token economics understate it.

\subsection{The Latency Condition}

Let $w = s_s/s_c$ be the ratio of strong to cheap generation throughput, so $w < 1$ when the cheap model generates faster. Collect the terms that do not scale with routing into a fixed overhead $\Omega = \left(\sum_i \ell_i - \ell_1\right) + H\tau$, comprising the extra prefills incurred by having $k$ stages instead of one plus the accumulated protocol tax. Normalising by baseline generation time gives $\theta = \Omega\,s_s/O_1$, the fixed overhead expressed as a fraction of the time the single call spends generating. The pipeline is faster when

\begin{equation}
\alpha\left[(1-f) + f w\right] + \theta < 1
\qquad\Longleftrightarrow\qquad
f \;>\; \frac{\alpha - 1 + \theta}{\alpha\,(1-w)},
\label{eq:latbreakeven}
\end{equation}

feasible only if $\alpha\,w + \theta < 1$, that is $\alpha < (1-\theta)/w$.

The latency condition has the same shape as the cost condition with one structural difference: the additive term $\theta$. The protocol tax has no cheap-model analogue -- routing work to a faster model shortens generation but does nothing to the envelopes, and a handoff costs what it costs regardless of who is on either side of it. The latency threshold is therefore strictly higher than the cost threshold whenever $\theta > 0$ and $w \ge r$, and the feasibility ceiling is correspondingly lower. This formalises an observation that practitioners report anecdotally: pipelines are brought back to cost parity by routing well before they are brought back to latency parity, and deep pipelines may never reach the latter at all.

\subsection{Separating Price from Volume}\label{sec:pricevolume}

Both conditions treat $\alpha$ and $r$ as independent, which they are not in practice. Substituting a cheap model changes token volume as well as price, and the volume change can go either way: a cheaper model may write less or reason more, and what it writes changes what the next stage is given to work with. Attributing the combined effect to price alone misstates what routing buys, because the volume component is a quality-relevant behavioural difference rather than a contractual discount -- it is invisible only while pass rates sit at ceiling. Operationally, $\alpha$ must be measured under the model assignment being evaluated, not under a uniform one, and an observed cost ratio between heterogeneous and uniform pipelines is best reported as the product of two factors: a \emph{volume factor}, the heterogeneous pipeline's own tokens priced at the strong model's rates relative to the uniform pipeline's cost, and a \emph{price factor}, its actual cost relative to that. Section~\ref{sec:results} reports both, and tests how well the model forecasts when $\alpha$ is taken from the uniform pipeline instead.

\subsection{Misrouting and Retries}

Both conditions assume the routing policy is correct. In deployment it is not, and a misroute -- the cheap model assigned to a step beyond its competence -- is paid for twice, once in the wasted cheap call and again in the retry on the strong model. Let $q$ be the probability that a routed step must be redone and $\kappa$ the cost of the retry relative to having used the strong model in the first place, with $\kappa > 1$ whenever the failed attempt is not refunded. The expected cost ratio becomes

\begin{equation}
\frac{\mathbb{E}[C_k]}{C_1} \;=\; \alpha\left[(1-f) + f\,r\,(1-q) + f\,q\,\kappa\right],
\label{eq:misroute}
\end{equation}

which tightens the threshold to $f > (\alpha-1)/[\alpha(1 - r + q(r - \kappa))]$ and lowers the feasibility ceiling accordingly. The practical content of \eqref{eq:misroute} is that the heterogeneity dividend degrades linearly in $q$ and vanishes entirely at $q = (1-r)/(\kappa - r)$: with a price ratio of one half and an unrefunded retry, a misroute rate of roughly one step in three erases the whole advantage. Because saturated benchmarks report $q = 0$ by construction, this term is invisible in exactly the regime where existing evaluations operate. The router condition of Section~\ref{sec:design}, reported in Section~\ref{sec:routing}, gives a direct estimate of how often the cheap model is sent work it then fails.

\section{Study Design}\label{sec:design}

\subsection{Conditions}

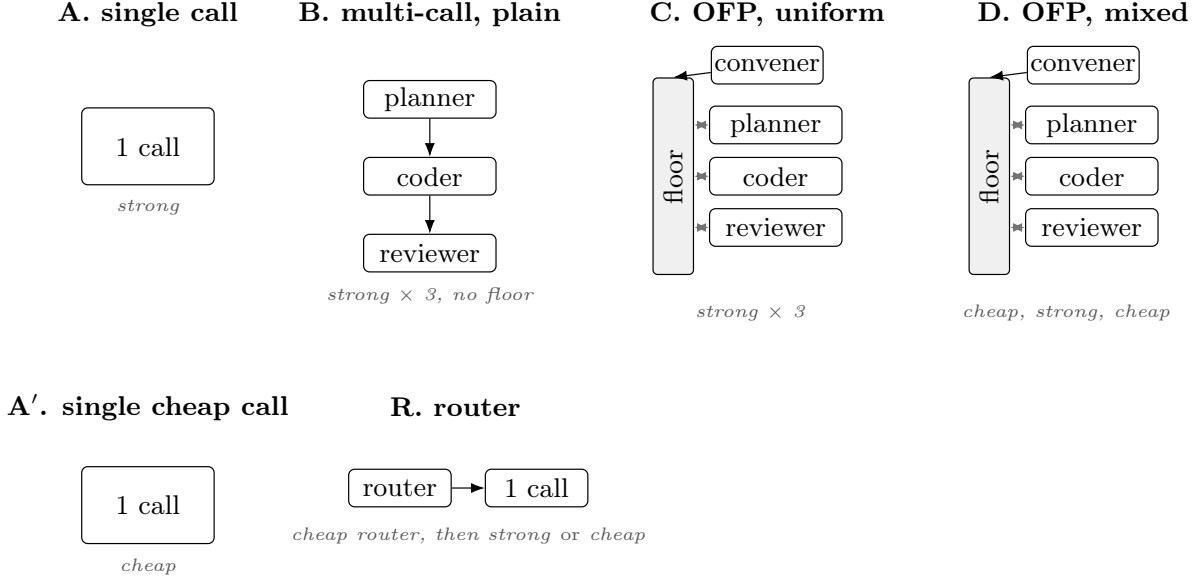
\begin{figure}[!htbp]
\centering
\resizebox{\textwidth}{!}{%
\begin{tikzpicture}[x=1cm, y=1cm]

\begin{scope}[xshift=0cm]
  \node[hdr] at (0,1.55) {A. single call};
  \node[callbox] (a1) at (0,0) {1 call};
  \node[mlab, below=1pt of a1] {strong};
\end{scope}

\begin{scope}[xshift=3.3cm]
  \node[hdr] at (0,1.55) {B. multi-call, plain};
  \node[agent] (b1) at (0,0.55) {planner};
  \node[agent] (b2) at (0,-0.35) {coder};
  \node[agent] (b3) at (0,-1.25) {reviewer};
  \draw[flow] (b1) -- (b2);
  \draw[flow] (b2) -- (b3);
  \node[mlab, below=1pt of b3] {strong $\times$ 3, no floor};
\end{scope}

\begin{scope}[xshift=7.0cm]
  \node[hdr] at (0.25,1.55) {C. OFP, uniform};
  \node[conv] (cv) at (0.25,0.95) {convener};
  \node[floorbar, minimum height=2.3cm] (cf) at (-0.85,-0.35) {\rotatebox{90}{floor}};
  \node[agent] (c1) at (0.35,0.25) {planner};
  \node[agent] (c2) at (0.35,-0.35) {coder};
  \node[agent] (c3) at (0.35,-0.95) {reviewer};
  \draw[flow] (cv) -- (cf.north);
  \draw[floorflow] (cf.east|-c1) -- (c1);
  \draw[floorflow] (cf.east|-c2) -- (c2);
  \draw[floorflow] (cf.east|-c3) -- (c3);
  \node[mlab] at (0.05,-1.95) {strong $\times$ 3};
\end{scope}

\begin{scope}[xshift=10.7cm]
  \node[hdr] at (0.25,1.55) {D. OFP, mixed};
  \node[conv] (dv) at (0.25,0.95) {convener};
  \node[floorbar, minimum height=2.3cm] (df) at (-0.85,-0.35) {\rotatebox{90}{floor}};
  \node[agent] (d1) at (0.35,0.25) {planner};
  \node[agent] (d2) at (0.35,-0.35) {coder};
  \node[agent] (d3) at (0.35,-0.95) {reviewer};
  \draw[flow] (dv) -- (df.north);
  \draw[floorflow] (df.east|-d1) -- (d1);
  \draw[floorflow] (df.east|-d2) -- (d2);
  \draw[floorflow] (df.east|-d3) -- (d3);
  \node[mlab] at (0.05,-1.95) {cheap, strong, cheap};
\end{scope}

\begin{scope}[yshift=-4.1cm]
  \node[hdr] at (0,1.05) {A$'$. single cheap call};
  \node[callbox] (e1) at (0,-0.1) {1 call};
  \node[mlab, below=1pt of e1] {cheap};

  \begin{scope}[xshift=3.3cm]
    \node[hdr] at (0.3,1.05) {R. router};
    \node[agent, minimum width=1.2cm] (r1) at (-0.35,0.1) {router};
    \node[agent, minimum width=1.2cm] (r2) at (1.25,0.1) {1 call};
    \draw[flow] (r1) -- (r2);
    \node[mlab] at (0.45,-0.45) {cheap router, then strong \emph{or} cheap};
  \end{scope}
\end{scope}

\end{tikzpicture}%
}
\caption{The six conditions. Top row: a single call on the strong model (A), and the three-role pipeline as plain calls (B), over OFP with the strong model in every role (C), and over OFP with the cheap model as planner and reviewer (D). Bottom row: the two baselines, a single call on the cheap model (A$'$) and a router (R). Only B and C differ in whether handoffs traverse the protocol; only C and D differ in model assignment.}\label{fig:conditions}
\end{figure}

Six conditions run each task under a different architecture (Fig.~\ref{fig:conditions}). A issues one call to the strong model and A$'$ one call to the cheap model. R is a router: the cheap model, at low reasoning effort, reads the task and answers SMALL or LARGE, and the chosen model then gets one call; any answer other than a clear SMALL goes to the strong model. B decomposes the task into three roles -- planner, coder, reviewer -- chained by direct API calls with no protocol layer. C runs the same three roles over OFP: each specialist is an independent BotAgent in its own process on its own port, exchanging Envelope, GrantFloorEvent, UtteranceEvent and YieldFloorEvent objects from the OFP Python package with a convener over HTTP. D is C with the cheap model as planner and reviewer and the strong model as coder. B, C and D use identical role prompts, and in all three the reviewer's output is the code that is graded, so in D the final code passes through the cheap model.

The design isolates one mechanism per contrast. B against A measures decomposition at a fixed model; C against B measures the protocol at fixed decomposition and model; D against C measures model assignment at a fixed architecture. A$'$ and R are the competitors any economic claim for D has to beat: A$'$ shows what the cheap model achieves alone, and R captures the heterogeneity dividend while paying neither the decomposition tax nor the protocol tax.

\subsection{Tasks}

Tasks come from two public, MIT-licensed benchmarks, in two tiers of difficulty. The \emph{function tier} draws on HumanEval \cite{humaneval2021}: one function per task, specified by its signature and docstring and graded by the task's own \texttt{check} function. The \emph{class tier} draws on ClassEval \cite{classeval2023}: one class per task with several methods sharing state, specified by a skeleton with docstrings and graded by the task's unittest classes. Decomposition has more to do in the class tier, where a plan has to cover several interacting methods.

Only ClassEval tasks whose solution and tests need nothing beyond the Python standard library are eligible (88 of 100). Every reference solution was run through the grader before use -- twice, or fifty times for tasks that use the random module -- and tasks that did not pass every time were dropped: three ClassEval tasks, whose tests depend on the current date, the machine's hostname and a random draw. The class tier is all \res{tasks:class} remaining tasks, so it involves no sampling choice; the function tier is \res{tasks:function} of HumanEval's 164 tasks, drawn with a fixed seed. Both downloads are pinned by SHA-256 hash. Each spec wraps the task in one instruction line before and one after, asking for the complete code and nothing else. Generated code is graded in a separate Python process, in a fresh temporary directory, with a 30-second timeout; a run counts as solved only if every test passes.

\subsection{Models, Prices and Reasoning}

The strong model is Claude Opus~5.5, at \$4/\$20 per million input/output tokens, and the cheap model Claude Haiku~5.5, at \$0.10/\$0.50,\footnote{Standard first-party API prices, \url{https://platform.claude.com/docs/en/about-claude/pricing}, accessed 15 October 2026. Haiku~5.5 is priced higher for prompts over 100{,}000 tokens; no prompt in this study comes close.} so $r_{\mathrm{in}} = r_{\mathrm{out}} = 0.025$. Every call is billed at its own model's list price, and the harness records input, output and reasoning tokens, cost and model-call time for each call. Both models reason before answering by default, and Opus~5.5 does not allow reasoning to be switched off; reasoning tokens are billed as output but never appear in the response. We therefore fix reasoning effort explicitly -- \texttt{medium} for every call except the router's, which is \texttt{low} -- and report reasoning tokens separately. Neither model accepts non-default sampling parameters, so repetitions sample each model's default stochasticity. Output is capped at 16{,}000 tokens per call, and no prompt caching is used.

\subsection{Measuring the Protocol}\label{sec:instrumentation}

The convener timestamps each handoff from just before it builds the request envelope to just after it has parsed the reply. The receiving agent reports how long its model call took, and the difference is the protocol's share of that handoff: envelope construction, JSON serialisation, the HTTP round trip, the agent's event dispatch and floor handling, and the reply envelope. All agents run as separate processes on one host and are started once for the whole run, so process start-up never falls inside a measured handoff. Loopback HTTP without TLS or authentication is the best case for any protocol, so the measurement is a lower bound on what a cross-host or cross-organisation deployment pays.

\subsection{Execution and Analysis}\label{sec:execution}

Each task is run \res{reps} times under each condition, \res{runs} runs in all. The run proceeds in rounds: each round holds one repetition of every (task, condition) pair in a fresh random order, so that drift in provider-side latency over the hours of the run cannot line up with any condition, and a spending cap fixed in advance can only ever cut the last round short. Runs are written as they complete, and an interrupted run resumes without repeating finished work. The run paused once when the account's API usage limit was reached and resumed without loss. In two runs a model request hung while the host was asleep and hit the client's 10-minute timeout before a retry succeeded; their wall time measured the host rather than the model, so both were discarded and re-run, and their \$0.07 is not included in the totals.

The unit of analysis is the task, not the run: repetitions of one prompt measure sampling noise, not generalisation. Each task's cost, wall time and solve rate are first averaged over its repetitions. Comparisons between conditions are paired by task. Ratios are ratios of mean per-task values, with 95\% confidence intervals from 5{,}000 bootstrap resamples of tasks, and $p$-values come from the Wilcoxon signed-rank test on per-task differences. pass@1 is the share of runs solved, and cost per solved task is total spend divided by the number of solved runs, following the cost-controlled practice of \cite{kapoor2024}.

The following hypotheses, and the criterion for each, were fixed before data collection and are implemented in the released analysis script; the repository history records them in commit \texttt{85a3748}, 25 minutes before the first run.
\begin{description}
\item[H1 (protocol).] The protocol accounts for under 1\% of the uniform OFP pipeline's median wall time, and the 95\% interval for C's cost relative to B's includes 1.
\item[H2 (model).] Holding token volumes at the uniform pipeline's per-role values and re-pricing them under D's model assignment forecasts D's cost relative to A to within 15\% of the observed ratio.
\item[H3 (economics).] The single cheap call A$'$ has a lower cost per solved task than the mixed pipeline D in both tiers.
\item[H4 (quality).] On class-level tasks, the OFP pipeline C solves more tasks than the single strong call A (paired by task, $p < 0.05$).
\end{description}

\section{Results}\label{sec:results}

\begin{figure}[!htbp]
\centering
\IfFileExists{figures/frontier.pdf}{\includegraphics[width=\textwidth]{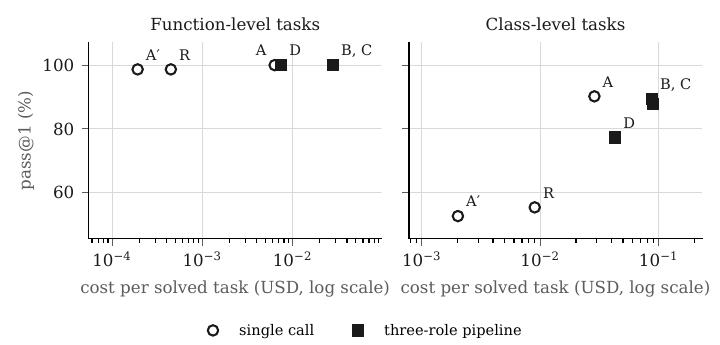}}{\fbox{\textcolor{red}{figure pending: run paper\_results.py}}}
\caption{Accuracy--cost frontier by tier. Each point is one condition: pass@1 against cost per solved task (log scale). Open circles are single calls (A, A$'$, R); filled squares are three-role pipelines (B, C, D). Conditions toward the upper left dominate those below and to their right.}\label{fig:frontier}
\end{figure}

The run took place between \res{date} and \res{dateend} and cost \$\res{spend} in total. Table~\ref{tab:main} gives pass@1, cost and tokens for every condition in each tier, Fig.~\ref{fig:frontier} the accuracy--cost frontier, and Table~\ref{tab:pairs} the paired comparisons over all \res{tasks} tasks. H1 was \res{H1}, H2 \res{H2}, H3 \res{H3} and H4 \res{H4}.

\begin{table}[!htbp]
\centering
\caption{Results by tier: \res{reps} runs per task and condition. pass@1 is the share of runs solved; cost per solved task is total spend divided by solved runs; cost and wall time are medians per run; token counts are means per run, and output includes the reasoning tokens shown in brackets.}\label{tab:main}
\setlength{\tabcolsep}{5pt}
\begin{tabular}{@{}lrrrrr@{}}
\toprule
Condition & pass@1 (\%) & \$/solved & Cost (\$) & Wall (s) & \shortstack[r]{Tokens in / out\\(reasoning)} \\
\midrule
\multicolumn{6}{l}{\emph{Function tier (HumanEval, \res{tasks:function} tasks)}} \\
\restable{main:function}
\midrule
\multicolumn{6}{l}{\emph{Class tier (ClassEval, \res{tasks:class} tasks)}} \\
\restable{main:class}
\bottomrule
\end{tabular}
\end{table}

\begin{table}[!htbp]
\centering
\caption{Paired comparisons over all \res{tasks} tasks. Cost and wall-time ratios are ratios of mean per-task values; $\Delta$pass@1 is the mean per-task difference in percentage points. Brackets give 95\% bootstrap intervals over tasks; $p$-values are from the Wilcoxon signed-rank test on per-task differences.}\label{tab:pairs}
\small
\setlength{\tabcolsep}{4pt}
\begin{tabular}{lcccccc}
\toprule
Comparison & Cost ratio & $p$ & Wall-time ratio & $p$ & $\Delta$pass@1 (pp) & $p$ \\
\midrule
\restable{pairs:all}
\bottomrule
\end{tabular}
\end{table}

\subsection{The Protocol's Cost}

Across \res{tau:turns} handoffs in C and D, the protocol took a median of \res{tau:median}\,ms per handoff (mean \res{tau:mean}\,ms, 95\% CI \res{tau:lo}--\res{tau:hi}\,ms; 95th percentile \res{tau:p95}\,ms): \res{tau:pipeline}\,ms per three-handoff pipeline, or \res{tau:shareC}\% of C's median wall time. A pipeline exchanged \res{envelope:kb}\,kB of envelopes on average. Measured end to end, the same quantity is much less precise: the mean per-task difference in wall time between C and B is \res{cb:diff}\,s (95\% CI \res{cb:lo} to \res{cb:hi}\,s, $p = \res{cb:p}$), an interval set by the variance of generation rather than by the protocol. C costs \res{costratio:all:C:B}$\times$ as much as B (Table~\ref{tab:pairs}).

\subsection{Accuracy and Cost}

Against the uniform pipeline C, the mixed pipeline D changes cost per task by a factor of \res{costratio:all:D:C} (95\% CI \res{costratiolo:all:D:C}--\res{costratiohi:all:D:C}) and pass@1 by \res{passdiff:all:D:C} percentage points; against the single strong call A, by a factor of \res{costratio:all:D:A} and \res{passdiff:all:D:A} points. The single cheap call A$'$ costs \res{costratio:all:Aprime:A}$\times$ as much as A per task, with a pass@1 difference of \res{passdiff:all:Aprime:A} points, and D costs \res{costratio:all:D:Aprime}$\times$ as much as A$'$. In the class tier, C's pass@1 differs from A's by \res{passdiff:class:C:A} points (95\% CI \res{passdifflo:class:C:A} to \res{passdiffhi:class:C:A}, $p = \res{passp:class:C:A}$). 

In the function tier every condition solves at least \res{pass:function:Aprime}\% of runs, so the conditions differ mainly in cost: A$'$ solves \res{pass:function:Aprime}\% at \$\res{cps:function:Aprime} per solved task, A solves \res{pass:function:A}\% at \$\res{cps:function:A}, and every pipeline costs more than A for no gain (Fig.~\ref{fig:frontier}). In the class tier the frontier runs from A$'$ through R to A. The cheap model alone solves \res{pass:class:Aprime}\% of class-level runs, and no condition solves more than A's \res{pass:class:A}\%. Every pipeline is dominated by A: B and C solve no more at about three times the cost, and D solves fewer at \res{costratio:class:D:A}$\times$ A's cost, its pass@1 differing from C's by \res{passdiff:class:D:C} points ($p = \res{passp:class:D:C}$). Because the reviewer's output is the graded artefact, the cheap model has the last word in D; the design does not separate how much of that loss comes from the cheap plan and how much from the cheap review.

\subsection{The Model's Quantities}

Decomposition at a fixed model inflates output volume by $\alpha_{\mathrm{out}} = \res{model:alphaout:all}$ and input volume by $\alpha_{\mathrm{in}} = \res{model:alphain:all}$ (function tier \res{model:alphaout:function} and \res{model:alphain:function}; class tier \res{model:alphaout:class} and \res{model:alphain:class}), against a feasibility ceiling of $1/r = \res{ceiling}$. D routes $f = \res{model:fout:all}$ of its output tokens to the cheap model, against a break-even threshold of $f^{\star} = \res{model:fstar:all}$. Holding token volumes at the uniform pipeline's values, \eqref{eq:full} forecasts D at \res{model:predfixed:all}$\times$ A (\res{model:predout:all}$\times$ from the output-only approximation \eqref{eq:breakeven}); the observed ratio of mean costs is \res{model:observed:all}$\times$, a forecast error of \res{model:forecasterr:all}\%. Under its own model assignment, D inflates output volume by \res{model:alphaoutD:all} and input volume by \res{model:alphainD:all} relative to A. Against C, D's cost ratio splits into a volume factor of \res{model:volume:all} and a price factor of \res{model:price:all}, and the coder, D's one strong-model stage, accounts for \res{model:codershare:all}\% of D's cost.

\subsection{Routing}\label{sec:routing}

R sent \res{router:small:function}\% of function-level runs and \res{router:small:class}\% of class-level runs to the cheap model, and the routing call itself accounted for \res{router:overhead:all}\% of R's cost. Of the runs routed to the cheap model, \res{router:smallfail:all}\% failed, an empirical estimate of the misroute rate $q$ in \eqref{eq:misroute}; of the runs routed to the strong model, \res{router:largewaste:all}\% concerned tasks the cheap single call solved in every repetition. An oracle that knew, per task, which single call solved it more often would reach a pass@1 of \res{oracle:pass:all}\% at \$\res{oracle:cps:all} per solved task, against \res{pass:all:R}\% at \$\res{cps:all:R} for R.

\subsection{Latency}\label{sec:latency}

Within each condition, output length and wall time correlate at $r$ between \res{corr:withinlo} and \res{corr:withinhi}; across all runs, at $r = \res{corr:all}$. In the single-call conditions Opus~5.5 produced about \res{tps:strong} output tokens per second of wall time and Haiku~5.5 about \res{tps:cheap}. D takes \res{wallratio:all:D:A}$\times$ as long as A per task and \res{wallratio:all:D:C}$\times$ as long as C.

\subsection{Failures}

Almost every unsolved run executed and failed at least one test. Across the six conditions, the number of runs that did not execute -- because they raised an error before their tests could run, crashed the grader or exceeded the 30-second limit -- was \ressum{compile}, of which \ressum{timeout} were timeouts; the number of runs that hit the 16{,}000-token output limit in any call was \ressum{truncated}; and the number of model calls declined by a safety classifier was \ressum{refusal}. Table~\ref{tab:failures} breaks down the unsolved runs by condition and cause.

\begin{table}[!htbp]
\centering
\caption{Unsolved runs by cause, over all tasks. ``Did not run'' includes timeouts, which are also counted separately. Truncated runs hit the 16{,}000-token output limit in at least one call; refusals count calls declined by a model's safety classifiers.}\label{tab:failures}
\small
\begin{tabular}{lccccc}
\toprule
Condition & Did not run & of which timeouts & Failed tests & Truncated & Refusals \\
\midrule
A & \res{fail:compile:A} & \res{fail:timeout:A} & \res{fail:tests:A} & \res{fail:truncated:A} & \res{fail:refusal:A} \\
A$'$ & \res{fail:compile:Aprime} & \res{fail:timeout:Aprime} & \res{fail:tests:Aprime} & \res{fail:truncated:Aprime} & \res{fail:refusal:Aprime} \\
R & \res{fail:compile:R} & \res{fail:timeout:R} & \res{fail:tests:R} & \res{fail:truncated:R} & \res{fail:refusal:R} \\
B & \res{fail:compile:B} & \res{fail:timeout:B} & \res{fail:tests:B} & \res{fail:truncated:B} & \res{fail:refusal:B} \\
C & \res{fail:compile:C} & \res{fail:timeout:C} & \res{fail:tests:C} & \res{fail:truncated:C} & \res{fail:refusal:C} \\
D & \res{fail:compile:D} & \res{fail:timeout:D} & \res{fail:tests:D} & \res{fail:truncated:D} & \res{fail:refusal:D} \\
\bottomrule
\end{tabular}
\end{table}

\section{Discussion}\label{sec:discussion}

\emph{The protocol is not where the money goes (H1).} At \res{tau:median}\,ms per handoff, three handoffs add about \res{tau:pipeline}\,ms to a pipeline whose model calls take tens of seconds, and the envelopes carry no billed tokens. Even a hundred times the measured figure -- a generous allowance for transport across hosts or organisations -- would add about two seconds to a pipeline that takes about twenty. An overhead this small is invisible end to end: the 95\% interval of the C-against-B comparison runs from \res{cb:lo} to \res{cb:hi}\,s. Arguments about the cost of interoperability protocols are therefore arguments about decomposition and model choice, not about the protocol.

\emph{$\alpha$ belongs to the model assignment, not the task (H2).} Re-pricing the uniform pipeline's token volumes under D's assignment forecast D at \res{model:predfixed:all}$\times$ A; D cost \res{model:observed:all}$\times$. The error lies entirely in volume. The price factor of \res{model:price:all} is close to the $(1-f) + fr \approx 0.28$ that the routed fraction implies, but the volume factor is \res{model:volume:all}: Haiku~5.5 reasons more than Opus~5.5 in the same roles (reasoning is \res{thinkshare:all:D}\% of D's output tokens against \res{thinkshare:all:C}\% of C's), and D's output volume relative to A rose from \res{model:alphaout:all} to \res{model:alphaoutD:all}. Anyone using \eqref{eq:breakeven} to plan a deployment has to measure $\alpha$ under the assignment being considered: with a cheap model that reasons at length, a decomposition that clears the threshold on paper -- here $f = \res{model:fout:all}$ against $f^{\star} = \res{model:fstar:all}$ -- can cost more than the single call it was meant to undercut.

\emph{The saving is in routing tasks, not roles (H3).} The single cheap call has the lowest cost per solved task in both tiers, as predicted, but cost per solved task is not the whole frontier: on class-level tasks A$'$ solves \res{pass:class:Aprime}\% of runs against A's \res{pass:class:A}\%. Per-role assignment works at the wrong granularity for these tasks. D pays strong-model prices for every task's coding stage and cheap-model quality on every task's plan and review, whereas the tasks themselves divide into those the cheap model solves alone and those it does not. An oracle choosing per task between the two single calls reaches \res{oracle:pass:all}\% at \$\res{oracle:cps:all} per solved task, half of A's \$\res{cps:all:A} at A's accuracy. Our zero-shot router captured the cost but not the accuracy: it sent \res{router:small:class}\% of class-level runs to the cheap model, and \res{router:smallfail:all}\% of all runs it routed there failed. Routing whole tasks needs no inter-agent protocol at all.

\emph{Decomposition bought no accuracy (H4).} On class-level tasks C's pass@1 differed from A's by \res{passdiff:class:C:A} points (95\% CI \res{passdifflo:class:C:A} to \res{passdiffhi:class:C:A}), at \res{costratio:all:C:A}$\times$ the cost and \res{wallratio:all:C:A}$\times$ the time. For a model as capable as Opus~5.5 on self-contained tasks of this size, a separate planner and reviewer add tokens without adding information the single call lacked. Whether that holds for under-specified or longer-horizon tasks, where a reviewer has real work to do, is the open question.

\emph{Implications for voice agents.} The two parts of the coordination tax weigh differently in a spoken conversation, where a pipeline's wall time is time the user spends waiting for a reply. The protocol's share, about \res{tau:pipeline}\,ms per three-handoff pipeline on one host, is far below what a listener would notice within a turn. Decomposition is not: the uniform pipeline took \res{wallratio:all:C:A}$\times$ as long as the single call, and moving two roles to the faster cheap model still left the mixed pipeline at \res{wallratio:all:D:A}$\times$. For voice agents the latency condition \eqref{eq:latbreakeven}, not the cost condition, is the binding one. A voice agent built on an interoperability protocol can therefore afford the protocol; whether it can afford a deep role decomposition behind every turn is a separate design decision, and the threshold and measurements above let a designer make it before deployment.

Whatever the cost results, they speak to only one half of the case for interoperability protocols. The other half rests on capabilities this study does not measure -- per-role backends, discovery, fault tolerance, audit trails and composition across organisational boundaries -- and those claims deserve measurement rather than assertion (Section~\ref{sec:future}).

\section{Threats to Validity}\label{sec:threats}

\emph{Task sample and contamination.} The \res{tasks} tasks come from two public benchmarks that the models have likely seen in training. Contamination inflates absolute pass rates for every condition alike; comparisons are paired by task, so it threatens the level of the results more than the differences between conditions. The two tiers support tier-level comparisons, not claims about narrower task types. Some benchmark tests also check behaviour the specification does not state: in ClassEval\_0, for example, the docstring promises a boolean while one test expects \texttt{None}, which the reference solution returns only by falling through. We use the benchmarks unchanged; such tests cap pass@1 for every condition alike rather than favouring one.

\emph{Repetitions are not tasks.} Repeated runs of one prompt measure sampling noise. All inference treats the task as the unit; the run count describes the measurement effort, not the sample size.

\emph{Provider-side variance.} Latency for hosted models depends on load conditions outside experimental control. Interleaved randomised execution prevents drift from aliasing onto conditions but does not make absolute latencies reproducible, and the measurements come from one run over two days, with the pauses described in Section~\ref{sec:execution}.

\emph{Price volatility.} The ratio $r$ is a commercial parameter that changes without notice. The model is stated in terms of $r$, so any specific price list is an instantiation of it rather than a finding.

\emph{Implementation dependence of $\tau$.} The measured protocol overhead belongs to one implementation -- the OFP Python package behind a Flask server -- on one host, not to the specification. It is a lower bound for deployments that cross hosts, networks or organisations.

\emph{Reasoning defaults.} Whether and how much a model reasons when a request does not say is a vendor default that differs between models and model versions. A comparison that leaves it unspecified compares defaults as well as models; we fix effort explicitly and report reasoning tokens per call.

\emph{The reviewer's output is graded.} In B, C and D the reviewer can repair the coder's mistakes or introduce new ones. This is a design choice that favours a strong reviewer; in D, the cheap model has the last word on every artefact.

{\sloppy\emph{The router.} R is a zero-shot prompt router. A trained router such as RouteLLM \cite{routellm2024} may route better; the oracle in Section~\ref{sec:routing} bounds what any router could achieve with these two models on these tasks.\par}

\emph{Provider monoculture and task family.} Both models come from one vendor, so protocol effects cannot be separated from vendor-specific serving behaviour, and all tasks are code generation, chosen because correctness is executable and therefore not judged by a model. Where quality must be assessed without execution, LLM-judge protocols carry known biases \cite{mtbench2023}.

\section{Future Work}\label{sec:future}

The study measures one protocol on one host, and four extensions follow directly. \emph{A second protocol}: running C and D over A2A as well would turn a single-protocol measurement into a comparative one and guard against attributing an implementation cost to protocol design. \emph{Distance}: repeating the protocol measurement across hosts and networks would replace the lower bound with deployment figures. \emph{Capabilities}: fault recovery (terminating an agent mid-task and measuring floor reassignment), fan-out over independent subtasks, the one regime in which a multi-agent architecture can beat a single call on wall time, and discovery (adding a capability through an Assistant Manifest rather than a redeployment) are the claims that motivate interoperability protocols and remain unquantified. \emph{Harder tasks}: a tier of deliberately under-specified requirements, a second model provider, a trained router, and a length-controlled ablation that re-runs the strong model under the cheap model's output budget would test whether the volume factor is paid for in quality.

\section{Reproducibility}\label{sec:integrity}

The harness, the task suite together with the builder and pinned dataset hashes that produced it, the raw run log, the two discarded runs, and the analysis script that generates every number, table and figure in Sections~\ref{sec:results} and~\ref{sec:discussion} are released under the MIT licence at \url{https://github.com/azettl/coordination-tax-harness}. The run log holds one row per run and, for every call within it, the model, input, output and reasoning tokens, cost, model-call time and protocol time, so that any reader can recompute the comparison, including in directions unfavourable to the protocol under test. A dry-run mode exercises the whole harness, including the OFP agent processes and the grader, without calling a model.

\section{Conclusion}

The question practitioners ask about agent interoperability protocols such as OFP or A2A -- are they slower and more expensive than just calling the model? -- has no answer as posed, because it holds fixed the one degree of freedom that decomposition buys. Reframed as an economic question, it becomes tractable: decomposition pays when the heterogeneity dividend $\alpha f (1-r)$ exceeds the decomposition tax $\alpha - 1$, which yields an explicit threshold on the routable fraction and a hard feasibility ceiling at $\alpha = 1/r$ beyond which no routing policy suffices. The latency version of the condition carries an additive protocol term with no cheap-model analogue, and is therefore strictly harder to satisfy. Because the model is stated in terms of $\alpha$, $f$ and $r$ rather than of any protocol's internals, it applies unchanged to A2A or any other inter-agent protocol; only the per-handoff tax $\tau$ has to be measured anew.

Measured on \res{tasks} code-generation tasks with OFP as the protocol under test, the costs separate cleanly. The protocol contributes \res{tau:median}\,ms per handoff and nothing to the bill. Decomposition contributes most of it -- about three times the single call's cost and time, with no gain in accuracy -- and moving roles to a cheap model recovers part of the cost while losing accuracy, because a cheap model that reasons at length raises the very token volume the threshold depends on. The saving that heterogeneous models offer is real, but on these tasks it belongs to choosing a model per task, and a protocol that makes per-role assignment easy, whether OFP, A2A or another, does not by itself capture it.

\section*{Acknowledgments}
We express our sincere appreciation to the Voiceinteroperability.ai \cite{ovoninter} Team (Linux Foundation AI \& Data Foundation) for their invaluable contributions and support in developing the Open-Floor-Protocol (OFP) Interoperable Standard, particularly to Deborah A. Dahl and Emmett Coin. Their expertise and suggestions have been pivotal in shaping the protocol on which this study is built. Both authors participate in the Open Voice Interoperability Initiative, which develops the Open Floor Protocol evaluated in this study.

	\bibliographystyle{plain}
	\bibliography{coordinationtax}

\end{document}